# Multi-probe study of excited states in $^{12}$C: disentangling the sources of monopole strength between the Hoyle state and $E_x = 13$ MeV

K. C. W. Li,[1,2,3,*] P. Adsley,[1,4,5] R. Neveling,[1] P. Papka,[1,2] F. D. Smit,[1] E. Nikolskii,[6,7] J. W. Brümmer,[1,2] L. M. Donaldson,[1,4] M. Freer,[8] M. N. Harakeh,[9] F. Nemulodi,[1] L. Pellegri,[1,4] V. Pesudo,[10,11] M. Wiedeking,[1,4] E. Z. Buthelezi,[1] V. Chudoba,[7] S. V. Förtsch,[1] P. Jones,[1] M. Kamil,[10] J. P. Mira,[1] G. G. O'Neill,[1,10] E. Sideras-Haddad,[4] B. Singh,[10] S. Siem,[3] G. F. Steyn,[1] J. A. Swartz,[1,2] I. T. Usman,[1,4] and J. J. van Zyl[2]

[1] *iThemba LABS, National Research Foundation, PO Box 722, Somerset West 7129, South Africa*
[2] *Department of Physics, University of Stellenbosch, Private Bag X1, 7602 Matieland, South Africa*
[3] *Department of Physics, University of Oslo, N-0316 Oslo, Norway*
[4] *School of Physics, University of the Witwatersrand, Johannesburg 2050, South Africa*
[5] *IPNO, Université Paris-Sud 11, CNRS/IN2P3, Orsay, France*
[6] *NRC Kurchatov Institute, Ru-123182 Moscow, Russia*
[7] *Flerov Laboratory of Nuclear Reactions, JINR, RU-141980 Dubna, Russia*
[8] *School of Physics and Astronomy, University of Birmingham, Edgbaston, Birmingham, B15 2TT, United Kingdom*
[9] *Nuclear Energy Group, ESRIG, University of Groningen, 9747 AA Groningen, The Netherlands*
[10] *Department of Physics and Astronomy, University of the Western Cape, P/B X17, Bellville 7535, South Africa*
[11] *Centro de Investigaciones Energéticas, Medioambientales y Tecnológicas, Madrid 28040, Spain*

**Background:** The Hoyle state is the archetypal $\alpha$-cluster state which mediates the $3\alpha$ reaction to produce $^{12}$C and is of great interest for both nuclear structure and astrophysics. Recent theoretical calculations predict a breathing-mode excitation of the Hoyle state at $E_x \approx 9$ MeV. Its observation is hindered by the presence of multiple broad states and potential interference effects. An analysis with Gaussian lineshapes of measurements at the Research Center for Nuclear Physics (Osaka University) with the Grand Raiden spectrometer suggested that additional strength was needed at $E_x \approx 9$ MeV to reproduce the data; this analysis did not account for the well-known threshold effects observed in $^{12}$C. Nevertheless, various theoretical studies have since concluded that this additional strength corresponds to the predicted breathing-mode excitation of the Hoyle state. To meaningfully identify a new source of monopole strength in this astrophysically significant region, a more appropriate phenomenological analysis which accounts for penetrability and interference effects must be used to determine whether the data can be explained with previously established states.

**Purpose:** We aim to investigate the monopole strength in the astrophysically important excitation-energy region of $^{12}$C between $E_x = 7$ and 13 MeV to determine whether the previously established sources of monopole strength are able to reproduce the data.

**Method:** The $^{12}$C$(\alpha, \alpha')^{12}$C and $^{14}$C$(p, t)^{12}$C reactions, which are expected to exhibit contrasting selectivity towards different monopole excitations, were employed at various detection angles and beam energies to populate states in $^{12}$C. The inclusive excitation-energy spectra were simultaneously analyzed with multilevel, multichannel lineshapes. Various scenarios with different sources of monopole strength and interference effects were considered to determine whether the ghost of the Hoyle state and the previously established broad $0_3^+$ state at $E_x \approx 10$ MeV are able to reproduce the observed monopole strength.

**Results:** Clear evidence was found for excess monopole strength at $E_x \approx 9$ MeV, particularly in the $^{12}$C$(\alpha, \alpha')^{12}$C reaction at 0°. This additional strength cannot be reproduced by the previously established monopole states between $E_x = 7$ and 13 MeV. Coincident charged-particle decay data suggest that the strength at $E_x \approx 9$ MeV is dominantly monopole, with no evidence of a $J > 0$ contribution.

**Conclusions:** The data support a new source of monopole strength at $E_x \approx 9$ MeV, which cannot be described with a phenomenological parametrization of all previously established states. An additional $0^+$ state at $E_x \approx 9$ MeV yielded a significantly improved fit of the data and is a clear candidate for the predicted breathing-mode excitation of the Hoyle state. Alternatively, the results may suggest that a more sophisticated, physically motivated parametrization of the astrophysically important monopole strengths in $^{12}$C is required.

## I. INTRODUCTION

The low-lying monopole strength of $^{12}$C remains an important research topic for both nuclear structure and astrophysics. Historically, there has been a strong focus on the $0_2^+$ Hoyle state located at $E_x = 7.65407(19)$ MeV, which is the archetypal $\alpha$-cluster state and which mediates the astrophysically significant $3\alpha$ reaction to produce $^{12}$C. Above the Hoyle state, the Evaluated Nuclear Structure Data File (ENSDF) database lists two $0^+$ resonances situated at $E_x = 9.930(30)$ and $10.3(3)$ MeV, with widths of $\Gamma = 2.710(80)$ and $3.0(7)$ MeV, respectively [1]. Due to the close proximity of these resonances with respect to their relatively large widths, it is currently understood that these two listed resonances are one and the same [2, 3], corresponding to a broad $0_3^+$ resonance at $E_x \approx 10$ MeV with $\Gamma \approx 3$ MeV. However, a more recent study of $\beta$-decay data from $^{12}$N and $^{12}$B indicates that the $0_3^+$ state may exhibit a higher resonance energy of $E_x \approx 11.2(3)$ MeV with a smaller width of $\Gamma \approx 1.5(6)$ MeV [4].

* k.c.w.li@fys.uio.no

The Hoyle state remains the focus of experimental and theoretical work with recent efforts to measure its properties such as its direct decay [5–10], gamma decay [11] and $E0$ decay branching ratios [12]. The total and partial widths of the Hoyle state are of great importance, both for testing our understanding of $\alpha$-particle clustering and the accurate modeling of stellar nucleosynthesis. The Hoyle state exhibits a narrow primary peak and a pronounced high-$E_x$ tail (the "ghost anomaly"), which results from the strong $\alpha$-cluster character of the Hoyle state and its proximity to the $\alpha$-separation energy [13]. This produces a strongly increasing $\alpha$-particle partial width for the Hoyle state in the excitation-energy region above the main peak, resulting in a distinctive high-energy component to the shape of the state. Currently, the total width of the Hoyle state is estimated to be $\Gamma = 9.3(9)$ eV, as deduced from the pair-decay partial width and branching ratio [13–15]. It has been suggested that the total width of the Hoyle state can also be indirectly measured through the shape of the ghost [16]. However, this is complicated by the limited knowledge of broad states between $E_x = 7$ and 13 MeV.

Theoretical studies have predicted an additional source of monopole strength resulting from the breathing-mode excitation of the Hoyle state which would lie at $E_x \approx$ 9 MeV with $\Gamma \approx 1.5$ MeV between the $0_2^+$ Hoyle state and the broad $0_3^+$ state at $E_x \approx 10$ MeV. Two independent calculations using the orthogonality condition model (OCM), which is suited for the study of states near and above the particle threshold energy, have predicted this additional collective $J^\pi = 0^+$ state at $E_x = 8.95$ MeV with $\Gamma = 1.48$ MeV [17, 18] and at $E_x = 8.09$ MeV with $\Gamma = 1.68$ MeV [19] with similar properties to the Hoyle state and may correspond to a higher nodal state of the Hoyle state [18]. A study with time-dependent fermionic molecular dynamics predicts two modes of collective isoscalar monopole excitations in $^{12}$C: one being a crossing of the $\alpha$ clusters through the center-of-mass of the system and the other a small-amplitude breathing-mode at lower excitation energies [20]. Generator coordinate method (GCM) calculations predict two distinct monopole excitations above the Hoyle state at $E_x = 9.38$ MeV and $E_x = 11.7$ MeV with different characters: the lower mode corresponds to a breathing-mode excitation of the Hoyle state and the higher mode corresponds to a bent-arm $3\alpha$ structure [21–23]. A variational calculation performed alongside the GCM calculation produced a large monopole transition strength between the Hoyle state and its predicted breathing-mode excitation [23]. Similarly, a calculation using the real-time evolution method predicts a dilute $2\hbar\omega$ breathing-mode excitation of the Hoyle state with a large associated monopole transition strength of 6.2 Weisskopf units [24].

Identification of this predicted breathing-mode excitation is complicated by theoretical and experimental factors. The high-energy tail of the Hoyle state extends some considerable energy above the main peak of the Hoyle state, overlying the region in which the breathing-mode is predicted. Another source of uncertainty in the broad strength at $E_x \approx 10$ MeV is the existence of the $2^+$ rotational excitation of the Hoyle state, which has been the subject of a decades-long search, culminating in its identification at $E_x = 9.870(60)$ MeV with $\Gamma = 850(85)$ keV through both inelastic scattering and photodisintegration [25–27]. This region was studied by Itoh *et al.* through the $^{12}$C$(\alpha,\alpha')^{12}$C reaction with $E_\alpha = 386$ MeV between $\theta_{\text{c.m.}} = 0°$ and $15°$. A peak-fitting analysis with Gaussian lineshapes of the resulting excitation-energy spectra required an additional peak at $E_x \approx 9.04(9)$ MeV with $\Gamma = 1.45(18)$ MeV to reproduce the data, however such Gaussian lineshapes do not capture the physical effects of near-threshold resonances or interference [25]. Since a multipole decomposition analysis (MDA) revealed the excitation-energy region at $E_x \approx 9$ MeV to be dominantly populated by monopole strength, a number of authors [19, 23, 28, 29] have discussed the additional Gaussian peak in the context of the breathing-mode excitation of the Hoyle state. However, a more detailed analysis is required, taking into account the complex shape of the Hoyle state and potential interference effects between the known resonances [30–35]. The objective of this work is to study the sources of monopole strength between $E_x = 7$ and 13 MeV and determine whether the data can be explained by the contributions of the two previously established sources of monopole strength and the associated interference effects. Disentangling these contributions to the monopole strength of $^{12}$C is important, both for understanding the nuclear structure of $^{12}$C and the $3\alpha$ reaction rate, which is dependent on the theoretical description of the Hoyle state as well as other additional sources of monopole strength. It is imperative that these factors are understood in order to provide a robust evaluation for the $3\alpha$ rate (which is beyond the scope of the present work). The main implications of this work are summarized in Ref. [36] and the details of the analysis are reported here.

## II. EXPERIMENTAL APPARATUS

Measurements of the $^{12}$C$(\alpha,\alpha')^{12}$C and $^{14}$C$(p,t)^{12}$C reactions at various laboratory angles and beam energies were performed at the iThemba Laboratory for Accelerator-Based Sciences (iThemba LABS) in South Africa. The experimental conditions are summarized in Table I. Proton and $\alpha$-particle beams were extracted from the Separated-Sector Cyclotron and transported down a dispersion-matched beamline to the target position of the K600 magnetic spectrometer [37]. The ejectiles were momentum-analyzed by the K600 spectrometer and detected at the focal plane of the K600 in a combination of vertical drift chambers (VDCs) and plastic scintillators (see Fig. 1). The experimental trigger was the signal from the plastic scintillator detectors. The horizontal and vertical position and trajectory of the ejectile are determined by the position measurements from the VDCs. For the measurements of $^{12}C(\alpha,\alpha')^{12}C$ at $\theta_{\text{lab}} = 0°$ ($E_{\text{beam}} = 200$ MeV) and $^{14}C(p,t)^{12}C$ experiment at $\theta_{\text{lab}} = 0°$, coincident charged-particle decay from excited $^{12}C$ states were detected in the CAKE, an array of double-sided silicon strip detectors [38]. The detection thresholds in the CAKE were approximately 600 keV for the $^{12}C(\alpha,\alpha')^{12}C$ measurements and 750 keV for the $^{14}C(p,t)^{12}C$. A comprehensive description of the experimental apparatus and techniques is reported elsewhere [39].

The enriched $^{14}C$ targets were produced through the thermal cracking of $CH_4$ gas (with $\approx 80\%$ enrichment of $^{14}C$) onto 100-$\mu$m-thick superheated Ta(Nb) film. The resulting 280-$\mu$m- and 300-$\mu$m-thick $^{14}C$ foils were then removed from the tantalum on a water surface [40].

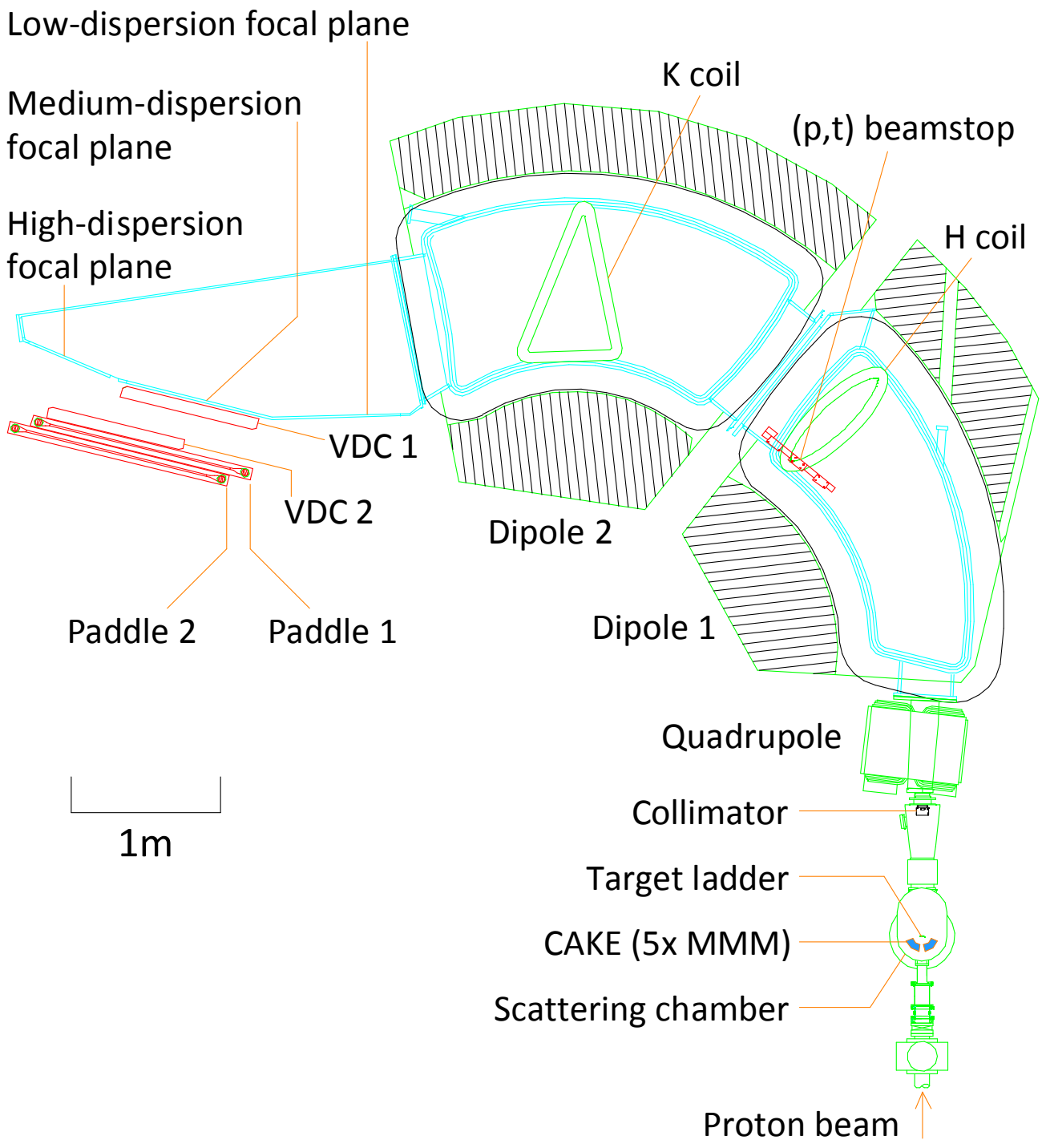

FIG. 1. Schematic top-view diagram of the K600 spectrometer positioned at $\theta_{\text{lab}} = 0°$, with the focal-plane detectors located in the medium-dispersion focal plane as used for the $^{14}C(p,t)^{12}C$ measurements of this work.

## III. DATA ANALYSIS

A simultaneous analysis of the inclusive excitation-energy spectra was performed with a fitting code which employed phenomenological multi-level, multi-channel lineshapes which include the energy dependence of the penetrability and interference effects, following the **R**-matrix formalism of Lane and Thomas [41]. The instrumental backgrounds were simultaneously fitted and account was taken for the experimental factors of each measurement. The formalism for the intrinsic nuclear lineshapes is presented in Section III A. The method of fitting the experimentally observed lineshapes, which are a function of the intrinsic lineshapes and experimental factors, is detailed in Section III B. The instrumental background and contaminants are detailed in Section III C.

The analysis of coincident charged-particle decays from excited $^{12}C$ states is presented in Section III D for the measurements of $^{12}C(\alpha,\alpha')^{12}C$ at $\theta_{\text{lab}} = 0°$ with $E_{\text{beam}} = 200$ MeV and $^{14}C(p,t)^{12}C$ at $\theta_{\text{lab}} = 0°$ with $E_{\text{beam}} = 100$ MeV. The angular correlations of decay are used to disentangle and constrain the contributions from different levels to the broad overlapping structures between $E_x = 7$ and 13 MeV.

### A. Intrinsic Lineshapes: Energy Dependence and Interference Effects

In contrast to the elastic resonance-scattering derivation of **R**-matrix theory, the measurements of this work all correspond to direct incoming (populating) channels for the target ($A$), projectile ($a$), recoil ($B$) and ejectile ($b$) nuclei of the form:

$$A + a \to B + b \qquad (B \to C + c), \tag{1}$$

where $C + c$ are the decay products of the recoil nucleus, which is modeled to proceed exclusively through two-body decay. As pioneered by Barker [42, 43], the cross section for resonances populated through direct reactions can be parameterized through a modification of the cross section for resonant scattering, as described in Ref. [41]. The intrinsic spectral lineshape for a reaction, which corresponds to the form of Equation 1, is expressed in terms of the level matrix $A$ as

$$N_{ab,c}(E) = P_c \left| \sum_{\lambda,\mu}^{N} G_{\lambda ab}^{\frac{1}{2}} \gamma_{\mu c} A_{\lambda\mu} \right|^2, \tag{2}$$

where $\gamma$ is the reduced-width amplitude. Subscript $ab$ denotes the $A + a \to B + b$ reaction channel and subscript $c$ denotes the $B \to C + c$ decay channel. $P_c$ is the penetrability and the incoming width has been replaced with a feeding factor, $G_{ab}$, which captures the excitation-energy dependence for the incoming reaction channel. The total width of the $\mu^{\text{th}}$ level is expressed as a sum over the decay-channel widths

$$\Gamma_\mu(E) = \sum_{c'} 2\gamma^2_{\mu c'} P_{c'}(\ell, E), \tag{3}$$

where $\gamma^2$ is the reduced width and $c'$ is a summation index over the decay channels. The penetrability for decay channel $c$, with an orbital angular momentum of the decay, $\ell$, is typically expressed as

$$P_c(\ell, E) = \frac{ka_c}{F_l(\eta, ka_c)^2 + G_l(\eta, ka_c)^2}, \tag{4}$$

where $F_l(\eta, ka_c)$ and $G_l(\eta, ka_c)$ are the regular and irregular Coulomb functions, respectively, $k$ is the wavenumber, $a_c$ is the fixed channel radius and $\eta$ is the dimensionless Sommerfeld parameter [44].

TABLE I. Summary of all experimental parameters. For the $^{12}$C($\alpha,\alpha'$)$^{12}$C and $^{14}$C($p,t$)$^{12}$C reactions, the focal-plane detector system was configured at the high- and medium-dispersion focal planes, respectively.

| Reaction | K600 angle ($\pm\theta$)[a] [deg] | $E_{\text{beam}}$ [MeV] | Target ($\mu$g/cm$^2$) | Accepted $E_x$ range[b] [MeV] | Fitted $E_x$ range [MeV] | Resolution FWHM[keV] | Ejectile energy loss[c] $c$ [keV] | $t$ [keV] |
|---|---|---|---|---|---|---|---|---|
| $^{12}$C($\alpha,\alpha'$)$^{12}$C | 0 (2.0) | 118 | $^{\text{nat}}$C (1053) | 5.0 - 14.8 | 5.0 - 14.8 | 50.9(2) | 7.31(10) | 89.6(7) |
| | 0 (1.91) | 160 | $^{\text{nat}}$C (300) | 7.3 - 20.7 | 7.3 - 20.0 | 48.1(2) | 2.12(4) | 93.00(5) |
| | 0 (2.0) | 200 | $^{\text{nat}}$C (290) | 9.7 - 25.2 | N.A.[d] | | | |
| | 10 (1.91) | 196 | $^{\text{nat}}$C (290) | 7.2 - 28.6 | 7.15 - 21.5 | 50.7(4) | 7.1(1) | 133.3(1) |
| $^{14}$C($p,t$)$^{12}$C | 0 (2.0) | 100 | $^{14}$C (280)[e] | 4.3 - 17.6 | 6.0 - 15.3 | 36.7(7) | 1.3(1) | |
| | 21 (1.91) | 67.5 | $^{14}$C (300)[f] | 7.2 - 14.5 | 6.8 - 14.5 | 34.2(6) | 0.29(7) | 73.6(4) |

[a] The collimator opening angle is shown in brackets.
[b] Determined by the range of excitation energies with full acceptance by the K600 spectrometer and the focal-plane detector system.
[c] Parameters $c$ and $t$, which approximate the target-related energy loss of the ejectile, correspond to the scale and location parameters of the Landau distribution, respectively.
[d] This measurement was excluded from the global fit analysis of inclusive focal-plane spectra. See Section IV for details.
[e] Enrichment of ≈87% $^{14}$C and ≈13% $^{12}$C.
[f] Enrichment of ≈80% $^{14}$C and ≈20% $^{12}$C.

The form of the penetrability in Equation 4 approximates the excitation energies of the product nuclei states to be infinitely narrow. For $\alpha$ decay from $^{12}$C to the ground and first-excited states of $^8$Be, respectively referred to as $\alpha_0$ and $\alpha_1$ decay, this approximation is not valid: the $0^+_1$ ground state exhibits a non-negligible high-energy tail which is analogous to the Hoyle state's ghost and the $2^+_1$ state at $E_x = 3.030(10)$ MeV has a width of $\Gamma = 1.513(15)$ MeV. In order to account for the finite widths of $^8$Be daughter states, a modified form of the penetrability is employed in this work:

$$\mathcal{P}_c(\ell, E) = \frac{\int_{E_0}^{E} P_c(\ell, E - E')\rho(E')\, dE'}{\int_{E_0}^{E} \rho(E')\, dE'}, \tag{5}$$

where $\rho(E')$ is the intrinsic lineshape of the populated $^8$Be state and $E_0$ is the minimum energy for the decay channel. For the inherently unbound $^8$Be nucleus, $E_0$ corresponds to the $\alpha$-$\alpha$ separation energy of $S_\alpha = -91.94$ keV. An alternative prescription for decay penetrabilities which proceed through broad intermediate states is given by the work of Lane and Thomas [41], expressed as

$$\mathcal{P}_c(\ell, E) = \frac{1}{\pi}\int_{E_0}^{E} P_c(\ell, E - E')\rho(E')\, dE', \tag{6}$$

with similar implementations used in Refs. [4, 45]. The results in the main body of this work correspond to the penetrability prescription of Equation 5. This choice is motivated by the principle that for the weighted average of the penetrability, where the probability distribution corresponds to the intrinsic lineshape of the daughter state, the normalization factor should correspond to the energetically accessible region of the daughter state. For completeness, the entire analysis of this work has been repeated using the penetrability prescription of Equation 6, with the results presented in the appendix.

To determine the modified penetrabilities with Equation 5, the intrinsic lineshapes for the $0^+_1$ and $2^+_1$ states of $^8$Be were determined with $\alpha$-$\alpha$ channel radii of $a_c = 6.5$ and 6.0 fm, respectively, corresponding to the study of $\alpha$-$\alpha$ elastic scattering, the $^9$Be($p,d$)$^8$Be reaction and $^8$Li $\beta$-decay by Barker *et al.* [43, 46]. Panel (b) of Fig. 2 presents the standard and modified $\alpha_0$ penetrabilities for $^{12}$C, which were determined with Equations 4 and 5, respectively. It was observed that by allowing $\alpha_0$ decay to populate the high-$E_x$ tail of the $0^+_1$ ground state of $^8$Be, the modified penetrabilities exhibit a slight suppression with respect to the standard penetrabilities. Panel (c) of Fig. 2 presents the standard and modified $\alpha_1$ penetrabilities and it is observed that accounting for the broad width of the $2^+_1$ state yields modified $\alpha_1$ penetrabilities which are non-zero below the $\alpha_1$ threshold for standard penetrabilities determined with Equation 4.

In order to achieve a self-consistent simultaneous analysis for all the measurements of this work, the feeding factors for each incoming channel were determined with CHUCK3, a coupled-channels nuclear reactions code [47]. For the analyzed inelastic $\alpha$-scattering measurements, it was assumed that the direct single-step mechanism dominates. The feeding factors for $J^\pi = 0^+, 1^-, 2^+, 3^-$ and $4^+$ states populated through the $^{12}C(\alpha,\alpha')^{12}C$ measurement at $\theta_{\rm lab} = 0°$ ($E_{\rm beam} = 160$ MeV) are presented in panel (a) of Fig. 3.

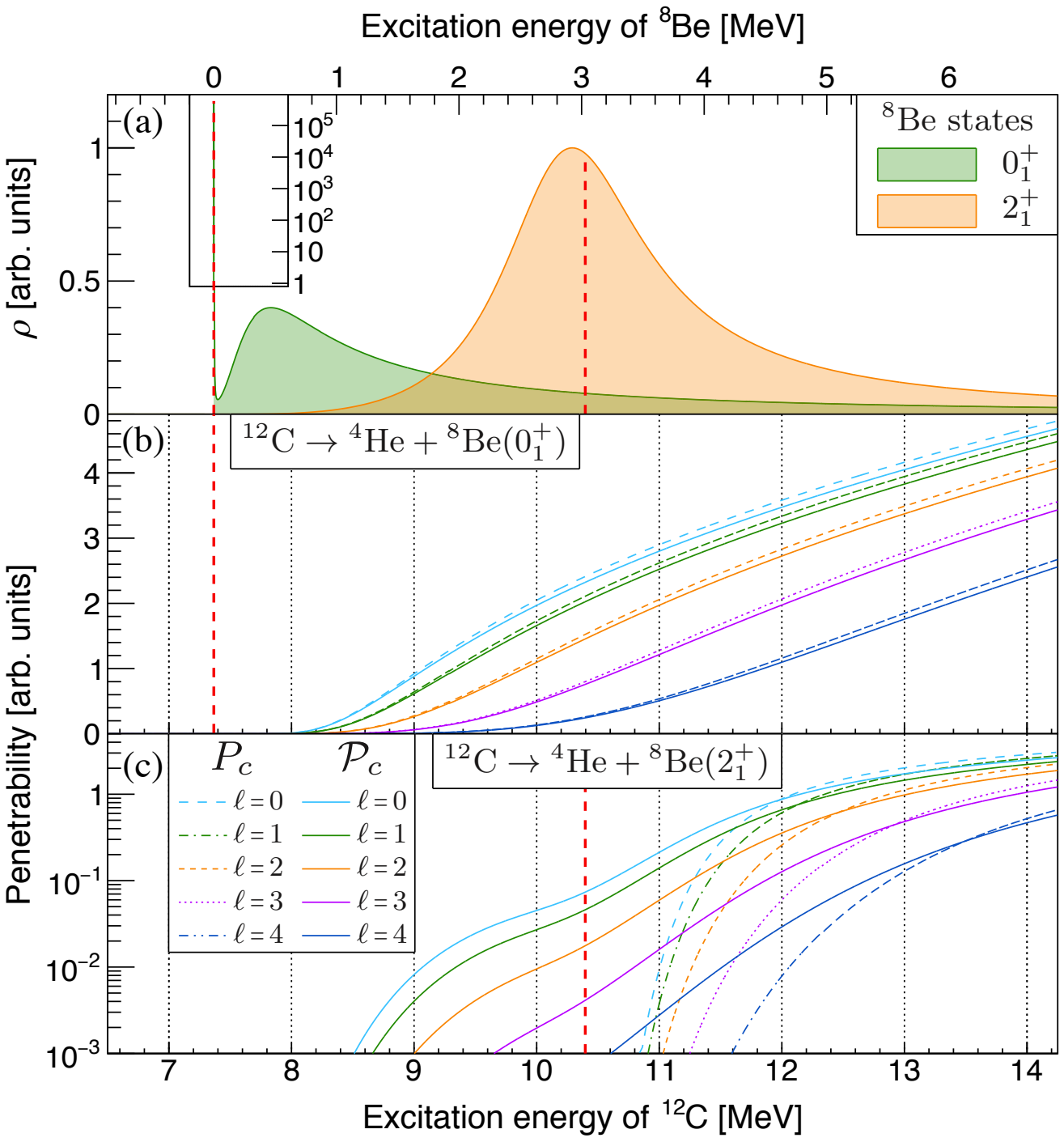

FIG. 2. (Color online) (a) The intrinsic lineshapes for the $0_1^+$ and $2_1^+$ states of $^8$Be, determined with an $\alpha$-$\alpha$ channel radius of $a_c = 6$ fm. The (b) $\alpha_0$ and (c) $\alpha_1$ penetrabilities for $^{12}$C, determined with a $^8$Be + $^4$He channel radius of $a_c = 6$ fm, and with Equation 4 (colored dashed and dash-dotted curves) or Equation 5 (colored solid curves). The red, vertical dashed lines indicate the resonance energies for the $0_1^+$ and $2_1^+$ states of $^8$Be, which respectively correspond to the $\alpha_0$ and $\alpha_1$ thresholds for standard penetrabilities calculated with Equation 4.

For the $^{14}C(p,t)^{12}C$ measurements, a coupled-channels calculation was performed to determine the feeding factors as the work of M. Yasue *et al.* indicated that the $^{14}C(p,t)^{12}C(t,t')^{12}C^*$ channel plays a significant role in reproducing the angular correlation of the Hoyle state [48]. The coupling scheme consisted of the $J^\pi = 0_1^+$, $2_1^+$, $0_2^+$ and $2_2^+$ states. The scheme is presented in Fig. 4 with the optimized intra- and inter-coupling parameters (see Table II for the associated spectroscopic amplitudes). The angular correlations and spectroscopic amplitudes for the $J^\pi = 0_1^+$, $2_1^+$, $0_2^+$ states from Ref. [48] were used to optimize the coupling parameters; an exception being the $2_2^+$ state, which has since been established to be located at $E_x = 9.870(60)$ MeV [1] and was assumed to belong to the rotational band with the Hoyle state being the band head. The feeding factors for the $J^\pi = 0_2^+$ and $2_2^+$ states are presented in panel (b) of Fig. 3. It was observed that the coupled-channels formalism affects the feeding factors with respect to a single-step mechanism.

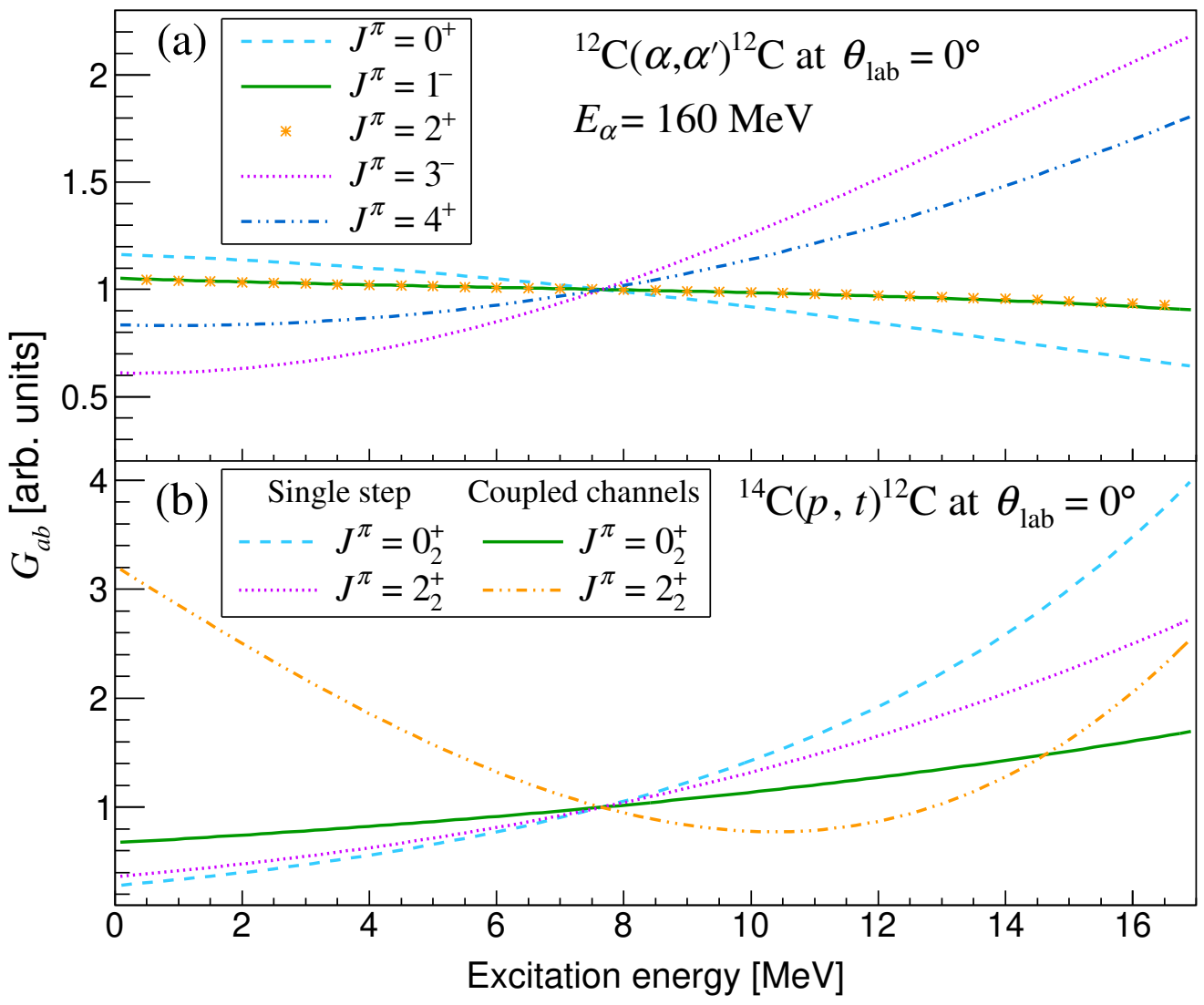

FIG. 3. The feeding factors, $G_{ab}$, for (a) the $^{12}C(\alpha,\alpha')^{12}C$ measurement at $\theta_{\rm lab} = 0°$ ($E_{\rm beam} = 160$ MeV) and (b) the $^{14}C(p,t)^{12}C$ measurement at $\theta_{\rm lab} = 0°$, arbitrarily normalised at $E_x = 7.654$ MeV.

TABLE II. The spectroscopic amplitudes corresponding to the $^{14}C(p,t)^{12}C$ coupling scheme presented in Fig. 4 (see Ref. [48]). Excitation energies are presented in units of MeV.

| $^{14}$C | | $^{12}$C | | Spectroscopic amplitudes | | |
|---|---|---|---|---|---|---|
| $E_x$ | $J^\pi$ | $E_x$ | $J^\pi$ | $(p_{1/2})^{-2}$ | $(p_{1/2}p_{3/2})^{-1}$ | $(p_{3/2})^{-2}$ |
| 0.0 | $0^+$ | 0.0 | $0^+$ | 0.7061 | | 0.4476 |
| | | 4.44 | $2^+$ | | 1.3061 | 0.5196 |
| | | 7.65 | $0^+$ | -0.6004 | | 0.3094 |
| | | 9.87 | $2^+$ | | -2.3661 | -2.8208 |

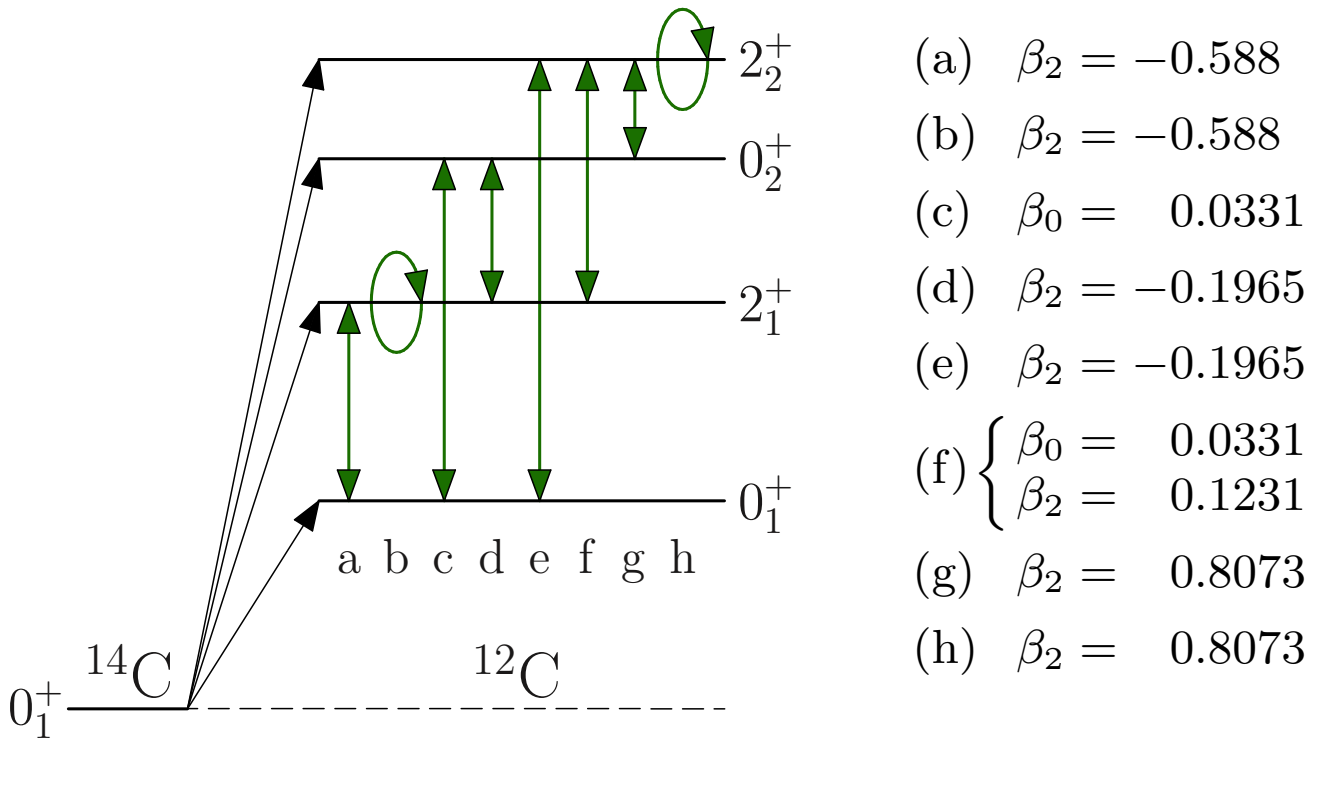

FIG. 4. The coupling scheme for the $^{14}C(p,t)^{12}C$ measurements analyzed in this work.

In the case of an isolated resonance, the corresponding lineshape corresponds to a single-level approximation [41] of the form

$$N_{ab,c}(E) = \frac{G_{ab}\,\Gamma_c}{(E - E_r - \Delta)^2 + \frac{1}{4}\Gamma^2}, \qquad (7)$$

where $E_r$ is the resonance energy and $\Delta \equiv \Delta_{11}$ is expressed as a sum over the decay channels with

$$\Delta_{\lambda\mu} = \sum_{c'} -(S_{c'} - B_{c'})\gamma_{\lambda c'}\gamma_{\mu c'}, \qquad (8)$$

where $S_c$ and $B_c$ are the shift factors and boundary condition parameters, respectively. For this work, the "natural" boundary condition, $B_c = S_c(E_r)$, was employed. The shift factors are typically expressed as

$$S_l(E) = \frac{ka_c\left[F_l(\eta, ka_c)F_l'(\eta, ka_c) + G_l(\eta, ka_c)G_l'(\eta, ka_c)\right]}{F_l(\eta, ka_c)^2 + G_l(\eta, ka_c)^2}, \qquad (9)$$

where $F_l'$ and $G_l'$ are the derivatives of the regular and irregular Coulomb functions, respectively. The shift factors exhibit the greatest energy dependence near particle thresholds and affect the intrinsic lineshape most strongly when the corresponding reduced widths are large—as is the case for states exhibiting significant $\alpha$ clustering such as the Hoyle state. Similar to the penetrabilities, the shift factors must also account for the finite widths of $^8$Be daughter states and consequently, a modified shift factor is employed for this work which uses a weighted averaging analogous to Equation 5.

For two overlapping levels of the same spin and parity, populated through a direct nuclear reaction (see Equation 1), the intrinsic lineshape is analyzed as a single entity through the two-level approximation [41]:

$$N_{ab,c}(E) = \sum_{s'l'} \frac{\left[\left(\overline{E}_2 - E\right) G_{1ab}^{\frac{1}{2}}\Gamma_{1c}^{\frac{1}{2}} + \left(\overline{E}_1 - E\right) G_{2ab}^{\frac{1}{2}}\Gamma_{2c}^{\frac{1}{2}} - \Delta_{12}\left(G_{1ab}^{\frac{1}{2}}\Gamma_{2c}^{\frac{1}{2}} + G_{2ab}^{\frac{1}{2}}\Gamma_{1c}^{\frac{1}{2}}\right)\right]^2 + \frac{1}{4}\left[\sum_{c'}\Pi_{c'ab}\Pi_{c'c}\right]^2}{\left[\left(\overline{E}_1 - E\right)\left(\overline{E}_2 - E\right) + \frac{1}{4}\left(\Gamma_{12}^2 - \Gamma_1\Gamma_2\right) - \Delta_{12}^2\right]^2 + \frac{1}{4}\left[\Gamma_1\left(\overline{E}_2 - E\right) + \Gamma_2\left(\overline{E}_1 - E\right) - 2\Delta_{12}\Gamma_{12}\right]^2}, \qquad (10)$$

which is a sum over the channel spins ($s'$) and orbital angular momenta ($\ell'$) of the decay channels, with

$$\Pi_{c'ab} = \Gamma_{1c'}^{\frac{1}{2}}G_{2ab}^{\frac{1}{2}} - \Gamma_{2c'}^{\frac{1}{2}}G_{1ab}^{\frac{1}{2}}, \qquad (11)$$

and

$$\Pi_{c'c} = \Gamma_{1c'}^{\frac{1}{2}}\Gamma_{2c}^{\frac{1}{2}} - \Gamma_{2c'}^{\frac{1}{2}}\Gamma_{1c}^{\frac{1}{2}}. \qquad (12)$$

The widths are defined as

$$\Gamma_{12} = \sum_{c'} 2P_{c'}\gamma_{1c'}\gamma_{2c'}, \qquad (13)$$

with the convention $\Gamma_\lambda \equiv \Gamma_{\lambda\lambda}$, and

$$\overline{E}_1 = E_1 + \Delta_1, \qquad (14)$$

with the convention $\Delta_\lambda \equiv \Delta_{\lambda\lambda}$. For a particular decay channel of two interfering resonances, it is observed that if the corresponding reduced-width amplitudes are of the opposite (same) sign, this results in constructive (destructive) interference in the excitation-energy region between the two resonance energies.

### 1. *Angular-momenta of decays*

Only the $\alpha_0$ and $\alpha_1$ decay modes are considered; the proton-decay channel is not open at the excitation energies considered in this work. For $\alpha_0$ decay, which can only occur from natural-parity states, both the channel spin and orbital angular momentum of decay are unique, thereby simplifying the form of Equation 10. For $\alpha_1$ decay, only the channel spin is invariably unique - for cases with multiple possible angular momenta of decay, the lowest $\ell$-value of decay is assumed to dominate and is set as the exclusive channel. This is reasonable as the penetrability decreases with increasing $\ell$-value. For unnatural-parity states in $^{12}$C, parity conservation dictates that $\alpha_0$ decay cannot occur, however $\alpha_1$ decay is possible. For the special case of monopole excitations, angular-momentum conservation dictates that $\alpha_0$ and $\alpha_1$ decay is exclusively $S$-wave ($\ell = 0$) and $D$-wave ($\ell = 2$), respectively.

### 2. *Channel radius and the Wigner limit*

For a particular decay channel $c$, the channel radius designates the boundary beyond which the potential between the two interacting nuclei can be approximated by the Coulomb potential and should therefore be larger than the summed radii of the two nuclei. The channel radius is parameterized as

$$a_c = r_0 \left( A_1^{1/3} + A_2^{1/3} \right) , \tag{15}$$

where $A_1^{1/3}$ and $A_2^{1/3}$ are the mass numbers and the value of $r_0$ is typically selected between 1.2 to 1.4 fm.

The Hoyle, $2_2^+$ and $0_3^+$ states of $^{12}$C are all understood to exhibit dilute densities with large radii [14]. Together with the $^4$He charge radius of 1.681(4) fm [49], the decay channels from such states require atypically large channel radii and for the Hoyle state in particular, the $\alpha_0$ channel radius is typically chosen as 6 fm [13]. The **R**-matrix analysis by S. Hyldegaard *et al.* employed a global channel-radius parametrization for states above (and including) the Hoyle state, with a discrete set of values: $r_0$ = 1.71, 2.09, 2.47, and 2.85 fm, corresponding to $a_c$ = 6.14, 7.50, 8.86 and 10.2 fm, respectively [32, 33]. For this work, an analogous global channel radius is implemented with test values from the discrete set of channel radii: $a_c$ = 6.0 to 11.0 fm in integer steps, corresponding to $r_0$ = 1.67, 1.95, 2.23, 2.51, 2.79 and 3.07 fm, respectively. Exceptions from this parametrization are the $3_1^-$ and shell-model-like $1_1^+$ states, for which the channel radius was chosen as $r_0$ = 1.3 fm, in accordance with a precise analysis of the $3_1^-$ state [50].

The Wigner limit, an upper limit on the reduced width (and thus on the corresponding partial width), is expressed in the seminal work of Wigner and Teichmann [51] as

$$\gamma_W^2 = \frac{3\hbar^2}{2\mu a_c^2}. \tag{16}$$

This limit is also alternatively formulated without the 3/2 factor [41, 52]. However, the larger form of Equation 16 is employed in this work for a conservatively larger parameter space. Unless explicitly stated, all fitted reduced-width parameters are constrained by the Wigner limit. In this work, both the Hoyle state and its $2_2^+$ rotational state are not constrained by the Wigner limit as both states exhibit widths which exhaust (or even exceed) the Wigner limit for certain choices of channel radii [13]. For the $i^{\text{th}}$ decay channel, the degree of clustering is indicated by the associated Wigner ratio:

$$\theta_i^2 = \frac{\gamma_i^2}{\gamma_{Wi}^2}, \tag{17}$$

whereby a value of $\gtrsim 0.1$ is understood to indicate significant clustering/preformation [53].

### B. Experimental Factors on the $E_x$ spectrum: Ion-optical corrections, VDC responses and target-related effects

The experimentally observed lineshape for a resonance is not only a function of the intrinsic lineshape (described in Section III A), but also of experimental factors. Careful consideration was given to the description of these effects to ensure a reliable extraction of $E_x$-dependent **R**-matrix parameters. A significant case that requires accurate parametrization is the fit for the primary peak of the Hoyle state: the extracted yield is directly linked to the strength of its ghost peak, which is submerged under the contributions from other broad states between $E_x \approx 8$ to 10 MeV.

The first experimental factor to take into account is the influence of the ion optics of the magnetic spectrometer on the observed focal-plane position spectra. Consider, for example, the kinematic broadening of two observed lineshapes on opposite sides of the focal-plane detector as shown in panel (a) of Fig. 5. While the K and H correction coils of the spectrometer make it possible to partially correct for kinematic broadening online, for this work more accurate corrections were subsequently performed during offline analysis, which also minimized the influence of the slight differential nonuniform response of the position-sensitive detectors on the lineshapes. The effect of kinematic broadening as well as higher-order aberrations were removed by iteratively correcting the horizontal focal-plane positions with measured parameters such as the vertical focal-plane position and the horizontal component of the scattering angles (both determined from raytracing), as well as the spectrometer time of flight. Such corrections can only be performed for a particular reaction, i.e. a given set of corrections for $^{12}\text{C}(\alpha,\alpha')^{12}\text{C}$ will still result in unfocused loci for the contaminant $^{16}\text{O}(\alpha,\alpha')^{16}\text{O}$ reaction (particularly at large values of $\theta_{\text{lab}}$). Having performed these corrections, the resulting shape of narrow states observed in the focal-plane position spectrum is determined by the beam parameters at the source point of the dispersion-matched system (comprising the beamline and magnetic spectrometer) as well as the ion optical characteristics of the combined system. The peak shapes are known to be well approximated by a Gaussian distribution, as confirmed through faint beam measurements [37] which allows for the direct observation of the dispersion-matched beam in the focal plane without any interactions with a target.

The second experimental factor results from the energy loss of the projectile and/or ejectile within the target, corresponding to a lower-than-expected measured rigidity for the ejectile, ultimately producing an artificial high-$E_x$ tail. Such energy loss effects can be parameterized with the Landau distribution [54]. Fig. 6 presents an analysis of these experimental factors for the $^{12}\text{C}(\alpha,\alpha')^{12}\text{C}$ measurement at $\theta_{\text{lab}} = 0°$ ($E_{\text{beam}}$ = 160 MeV), focused on (a) the $J^\pi = 2_1^+$ and (b) $J^\pi = 0_2^+$ states of $^{12}$C. The lineshape corresponding to a convolu-tion between the intrinsic lineshape and a Gaussian distribution accounts for the VDC response but results in a poor fit which cannot account for the observed asymmetry of the peak. Given the narrow width of the intrinsic lineshape for the $J^\pi = 2_1^+$ state and its location below particle threshold, the asymmetric high-$E_x$ tail on the observed lineshape is determined by experimental factors such as target-related energy loss. An additional convolution with a Landau distribution yields a significantly improved fit for $E_x$ values below the peak, however the high-$E_x$ tail is overestimated by $\approx 1\%$ of the peak maximum. Possible sources for this difference are imperfect kinematic corrections (which are limited by the experimental resolution) and/or scattering effects occurring within the exit window of the spectrometer. To account for this difference, a free truncation parameter ($t$ in Table I) was introduced and the resulting convolution yields the best fit and an accurate description for the total experimental response. An analogous observed-lineshape analysis for the $J^\pi = 0_2^+$ state of $^{12}$C is presented in panel (b). For each measurement, the optimized parameters for the VDC resolution and target-related energy loss were determined and assumed to be constant across the entire associated focal-plane spectrum; a valid approximation given the relatively small range of ejectile energies accepted by the K600 spectrometer.

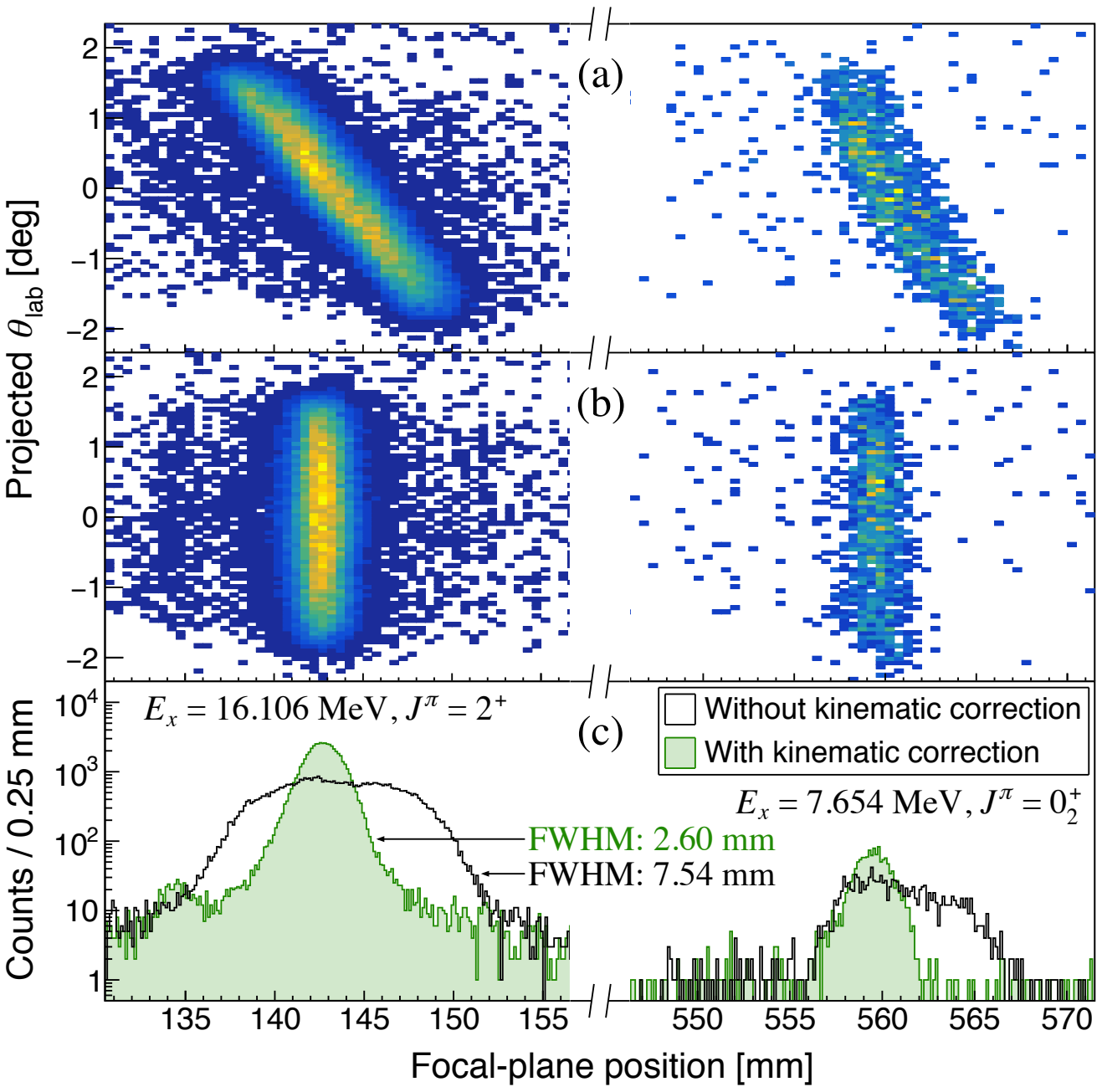

FIG. 5. Excited states from the $^{14}$C$(p,t)^{12}$C reaction at $\theta_{\rm lab} = 0°$ observed on opposite sides of the focal-plane detector. In panel (a) the effect of kinematic broadening as well as small contributions of higher-order aberrations can be seen, which are corrected in panel (b). The influence on the resulting lineshapes are shown in panel (c).

In summary, the experimental response was determined by accounting for the ion-optics of the magnetic spectrometer, the VDC response and the target-related energy loss. This was required for an accurate analysis of the focal-plane spectra (see Ref. [39] for details).

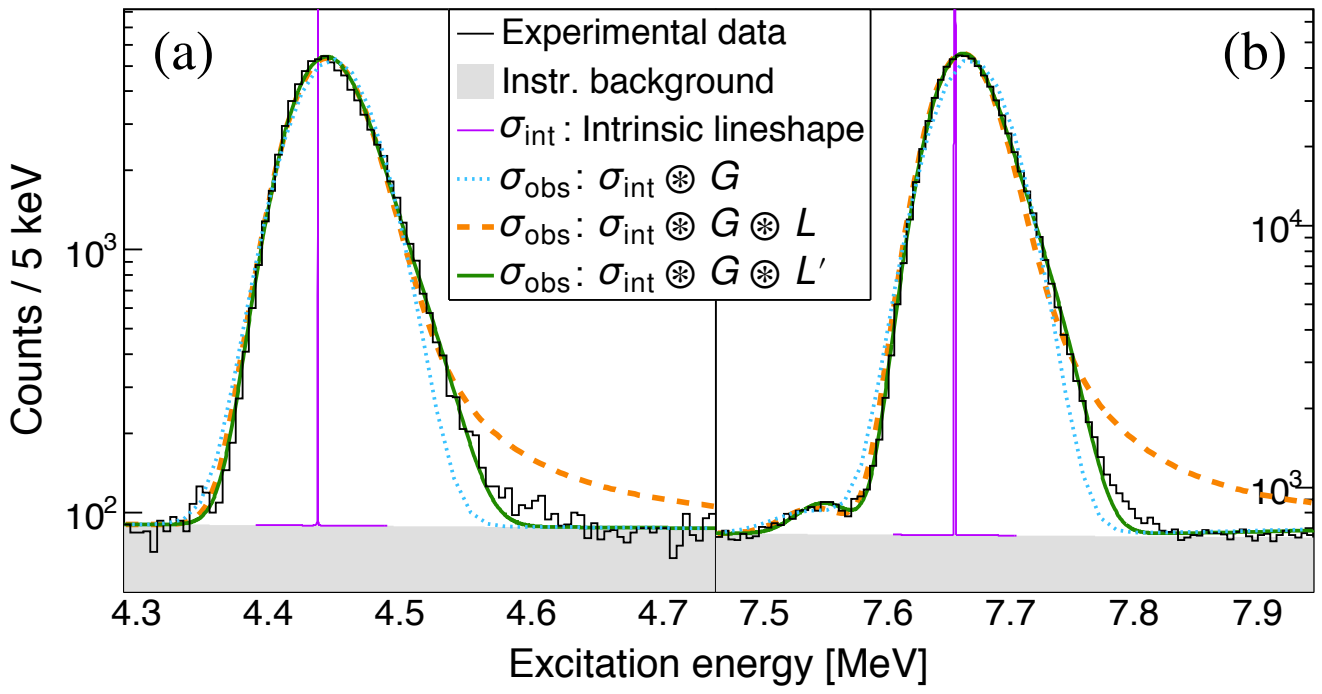

FIG. 6. The excitation-energy spectra from the $^{12}$C$(\alpha,\alpha')^{12}$C measurement at $\theta_{\rm lab} = 0°$ ($E_{\rm beam} = 160$ MeV), focused on modeling the VDC response and target-related attenuation for (a) the $J^\pi = 2_1^+$ and (b) $J^\pi = 0_2^+$ states of $^{12}$C. The spectrum presented in panel (a) corresponds to a subset of the data around the $2_1^+$ state. The small contaminant peak below the Hoyle state corresponds to the $E_x = 7.547(3)$ MeV state of $^{13}$C from the $^{13}$C$(\alpha,\alpha')$ reaction. The experimentally observed lineshapes ($\sigma_{\rm obs}$) are produced through circular convolution (denoted $\circledast$) of the intrinsic lineshape with a combination of Gaussian ($G$), Landau ($\ell$) and truncated Landau ($L'$) distributions.

### C. Instrumental background and contaminants

The dominant source of instrumental background within a focal-plane spectrum (particularly at $\theta_{\rm lab} = 0°$) results from small-angle elastic scattering off the target foil that is followed by re-scattering off any exposed part inside the spectrometer. These background events form smooth, slow-varying continuums and can be well characterized by operating the spectrometer in focus mode, whereby the quadrupole at the entrance to the spectrometer is used to vertically focus reaction products to a narrow horizontal band on the focal plane. A gate on this vertically focused band corresponds to the focal-plane spectrum for the target nucleus of interest and conversely, a gate on events which fall outside this focused band can be used to generate a sample of the instrumental background [37, 55]. Only for the measurements of $^{12}$C$(\alpha,\alpha')^{12}$C at $\theta_{\rm lab} = 10°$ and $^{14}$C$(p,t)^{12}$C at $\theta_{\rm lab} = 21°$ of this work, such instrumental background spectra could not be generated: the former was not performed in focus mode and the latter was performed with two VDCs that did not provide vertical-position information. However, these limitations are mitigated by the fact that measurements at non-zero angles typically exhibit low experimental backgrounds. Moreover, since the excitation-energy spectrum of $^{12}$C is devoid of strength between the $J^\pi = 2_1^+$ and Hoyle states, the low yields observed in this excitation-energy region for these ex-periments enabled the instrumental backgrounds to be reliably evaluated. The focal-plane spectra of interest and the associated instrumental background spectra are displayed in Fig. 7. In this analysis, the instrumental background spectra are not directly subtracted from the corresponding spectra of interest. Instead, all spectra were simultaneously fitted: each instrumental background spectrum was parameterized with a polynomial, which was then added with a free scaling parameter to the expected yield for the corresponding focal-plane spectrum of interest. This scaling is correlated with the relative size of the aforementioned vertical-position gates.

The observed contaminants are summarized in Table III and labeled in Fig. 7. For the $^{12}$C$(\alpha,\alpha')^{12}$C measurements at $\theta_{\mathrm{lab}} = 0^\circ$ and $10^\circ$, all contaminant states corresponded to inelastic alpha-scattering off $^{13}$C and $^{16}$O. No contaminating states from these nuclei were observed in the excitation-energy region of $E_x = 7.654$ to 9.2 MeV. Such potential contaminant states from $^{13}$C and $^{16}$O in the associated rigidity range are either narrow or in an excitation region with weak population of $^{12}$C excitations (just above the Hoyle state at $E_x \approx 8$ MeV) and would therefore be easily identified. Consequently, it is determined that there is negligible contribution from contaminating states in this excitation energy range. For the $^{16}$O state at $E_x = 9.585$ MeV, the sufficiently narrow width of $\Gamma = 420(20)$ keV should permit a significant contribution to be clearly identified. No significant contamination from this state was observed and this in agreement with the analogous measurement of $^{16}$O$(\alpha,\alpha')^{16}$O (at $\theta_{\mathrm{lab}} = 0^\circ$ with $E_\alpha = 200$ MeV) [56, 57] where the yield for this state was an order of magnitude less than that of the $^{16}$O state at $E_x = 12.049$ MeV, labeled (h). Using this yield ratio as an approximation, the expected contributions of the $E_x = 9.585$ MeV state (superimposed on the surrounding broad strength) are presented as violet lineshapes ($\varepsilon$). Under the assumption that this approximation is valid, it is observed that the expected contamination from the $E_x = 9.585$ MeV state in $^{16}$O is negligible.

For the $^{12}$C$(\alpha,\alpha')^{12}$C measurement at $\theta_{\mathrm{lab}} = 10^\circ$, significant contamination from elastic-scattering off hydrogen was observed at excitation-energies beyond $E_x \approx 15$ MeV. The large range of contamination results from the kinematic variation between the $^{1}$H$(\alpha,\alpha)$ and $^{12}$C$(\alpha,\alpha)^{12}$C reactions, which enabled the hydrogen-contamination events to be isolated (i) and removed from the focal-plane spectrum of interest.

For the $^{14}$C$(p,t)^{12}$C measurements, the self-supporting $^{14}$C-enriched targets exhibited additional contamination with respect to the $^{\mathrm{nat}}$C targets used for the $^{12}$C$(\alpha,\alpha')^{12}$C measurements. This is due to the entirely different manufacturing methods between the $^{\mathrm{nat}}$C and exotic $^{14}$C targets, in which the latter involved contact with iron, copper, tantalum and niobium. In addition to contamination from $^{13}$C and $^{16,18}$O, additional contaminant states from $^{56}$Fe, $^{63,65}$Cu and $^{14,15}$N were observed. Fortunately, all the identified contaminant nuclei only exhibit narrow states over the rigidity range between $E_x = 4$ and 16 MeV. The $^{14}$C targets used for both $^{14}$C$(p,t)^{12}$C measurements are from the same batch and are therefore expected to contain the same contamination. Consequently, it is suggested that the vastly different number of contaminant states observed between the $^{14}$C$(p,t)^{12}$C measurements at $\theta_{\mathrm{lab}} = 0^\circ$ and $\theta_{\mathrm{lab}} = 21^\circ$ are not only due to the different excitation-energy range measured, but also a relative decrease in cross section for the contaminant states at $\theta_{\mathrm{lab}} = 21^\circ$.

## D. Coincident charged-particle decay

The detection of coincident charged-particle decays from excited $^{12}$C states using the CAKE can help disentangle the contributions from different decay channels. Decaying particles may be identified by using the time-of-flight and the detected energy in the silicon detectors (see Ref. [38]). The matrices of particle ($\alpha$ or proton) en-ergy versus the excitation energy of $^{12}$C are presented in Fig. 8 for the measurements of $^{12}$C$(\alpha,\alpha')^{12}$C at $\theta_{\rm lab} = 0°$ ($E_\alpha = 200$ MeV) and $^{14}$C$(p,t)^{12}$C at $\theta_{\rm lab} = 0°$. The $\alpha_0$ and proton decays appear as well-defined loci and are labeled in the figure. In contrast, the indistinct locus between the $\alpha_0$ and $p_0$ decay modes corresponds to $\alpha$ particles emitted following the $\alpha$ decay of $^{12}$C (either to the broad $J^\pi = 2_1^+$ state of $^8$Be or the ghost of the $^8$Be ground state) and the subsequent $2\alpha$ breakup of $^8$Be.

TABLE III. Summary of observed contaminant states. The labels correspond to Fig. 7.

| Reaction | Angle [deg] | $E_{\mathrm{beam}}$ [MeV] | Label | Recoil $E_x$ [MeV] | Contaminant reaction |
|---|---|---|---|---|---|
| $^{12}$C$(\alpha,\alpha')$ | 0 | 118 | a | 6.049 | $^{16}$O$(\alpha,\alpha')$ |
| | | | b | 6.130 | $^{16}$O$(\alpha,\alpha')$ |
| | | | c | 6.917 | $^{16}$O$(\alpha,\alpha')$ |
| | | | d | 7.117 | $^{16}$O$(\alpha,\alpha')$ |
| | | | e | 7.547 | $^{13}$C$(\alpha,\alpha')$ |
| | | | f | 11.080 | $^{13}$C$(\alpha,\alpha')$ |
| | | | g | 11.520 | $^{16}$O$(\alpha,\alpha')$ |
| | | | h | 12.049 | $^{16}$O$(\alpha,\alpha')$ |
| $^{12}$C$(\alpha,\alpha')$ | 0 | 160 | d | 7.117 | $^{16}$O$(\alpha,\alpha')$ |
| | | | e | 7.547 | $^{13}$C$(\alpha,\alpha')$ |
| | | | f | 11.080 | $^{13}$C$(\alpha,\alpha')$ |
| | | | g | 11.520 | $^{16}$O$(\alpha,\alpha')$ |
| | | | h | 12.049 | $^{16}$O$(\alpha,\alpha')$ |
| $^{12}$C$(\alpha,\alpha')$ | 10 | 196 | a | 6.049 | $^{16}$O$(\alpha,\alpha')$ |
| | | | b | 6.917 | $^{16}$O$(\alpha,\alpha')$ |
| | | | c | 7.117 | $^{16}$O$(\alpha,\alpha')$ |
| | | | d | 7.547 | $^{13}$C$(\alpha,\alpha')$ |
| | | | i | 0.0 | $^{1}$H$(\alpha,\alpha)$ |
| $^{14}$C$(p,t)^{12}$C | 0 | 100 | j | 0.0 | $^{56}$Fe$(p,t)^{54}$Fe |
| | | | k | 2.832 ⋮ 3.264 | $^{65}$Cu$(p,t)^{63}$Cu |
| | | | l | 8.872 | $^{15}$N$(p,t)^{13}$N |
| | | | m | 2.949 | $^{56}$Fe$(p,t)^{54}$Fe |
| | | | n | 4.756 | $^{63}$Cu$(p,t)^{61}$Cu |
| | | | o | 0.0 | $^{13}$C$(p,t)^{11}$C |
| | | | p | 2.000 | $^{13}$C$(p,t)^{11}$C |
| | | | q | 0.0 | $^{16}$O$(p,t)^{14}$O |
| | | | r | 0.0 | $^{14}$N$(p,t)^{12}$N |
| | | | s | 0.961 | $^{14}$N$(p,t)^{12}$N |
| $^{14}$C$(p,t)^{12}$C | 21 | 67.5 | o | 0.0 | $^{13}$C$(p,t)^{11}$C |

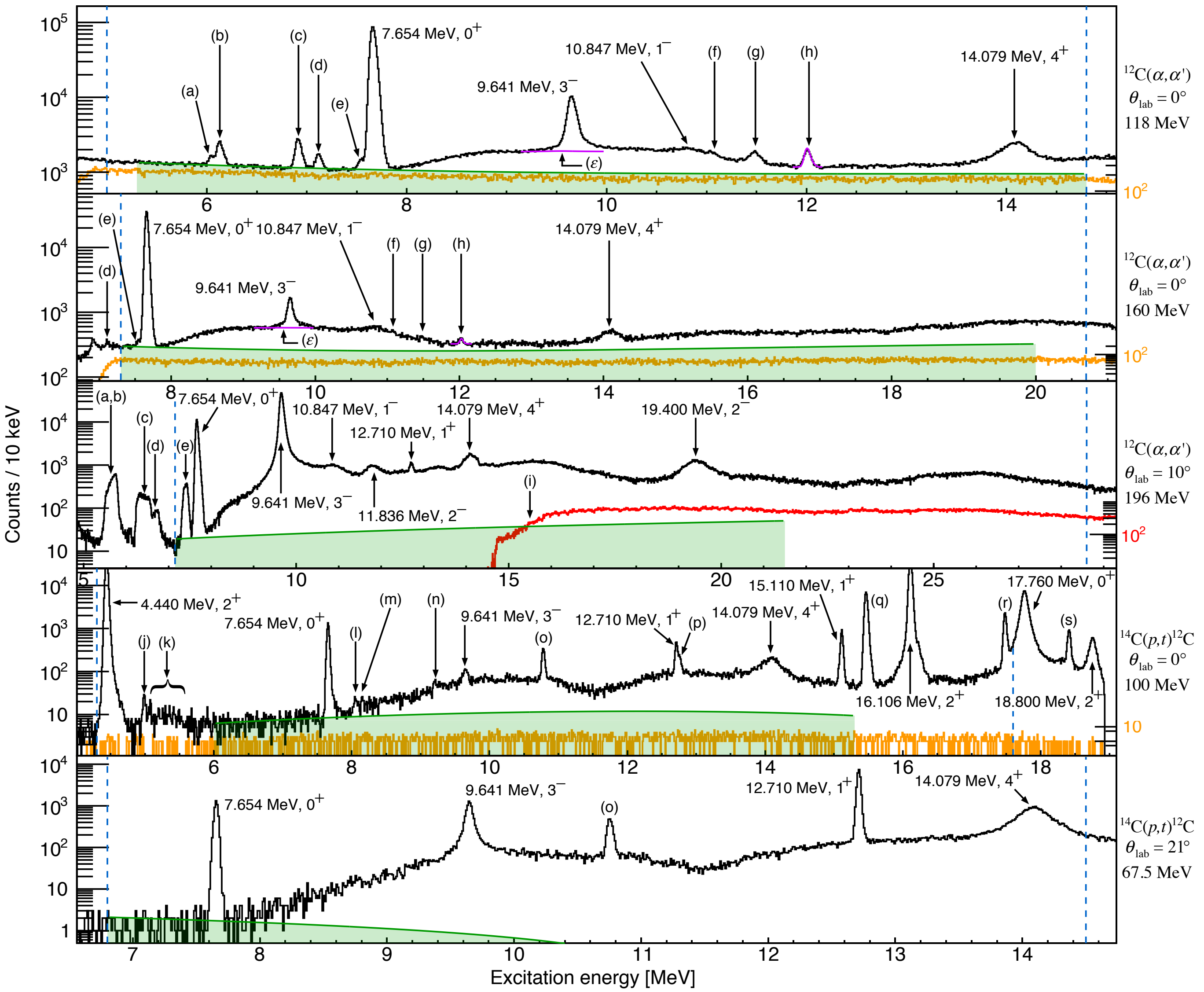

FIG. 7. Full-range excitation-energy spectra of all analyzed inclusive measurements. The instrumental background spectra and the associated scaled contributions to the spectra of interest are displayed in orange and green, respectively. The excitation energies, spins and parities for well-resolved $^{12}$C states are indicated. Contaminant peaks are labeled according to Table III. The red histogram (i) corresponds to contaminant events from the $^{1}$H$(\alpha,\alpha)$ reaction. The pair of blue, vertical dashed lines on each spectrum indicates the excitation-energy range of full acceptance by the spectrometer, as summarized in Table I. For details on the violet lineshapes denoted by $(\varepsilon)$ and (h), see Section III C.

Angular correlations of decay particles enable the assignment of spin and parity. In particular, $\alpha_0$ decay from $^{12}$C permits unambiguous assignments as it can only occur from natural-parity states with unique orbital angular momenta. The identification of different multipolarities within the broad overlapping structures between $E_x = 7$ and 13 MeV guides what multipolarities must be present to create a consistent model. The relative strengths of identified multipolarities enables the rejection of models if the associated fits result in inconsistent population strengths. Charged-particle decay at lower values of $\theta_{\rm lab}$ in the backward hemisphere exhibits more target-related energy loss and is more susceptible to be affected by the electronic thresholds of the CAKE. Consequently, detector channels situated at lower values of $\theta_{\rm lab}$ were ignored for affected decay modes.

Whilst both the $^{12}$C$(\alpha,\alpha')^{12}$C and $^{14}$C$(p,t)^{12}$C mea-

surements exhibited target contamination with $^{16}$O, only the $^{12}$C$(\alpha,\alpha')^{12}$C measurement yielded $^{16}$O contaminant states on the matrix of particle energy versus the excitation energy of $^{12}$C. This is due to the substantially different $Q$-values for the $^{14}$C$(p,t)^{12}$C and $^{16}$O$(p,t)^{14}$O reactions of $Q = -4.641$ and $-20.40562$ MeV, respectively. To determine the best quality models to describe the angular correlations, either the reduced chi-square statistic ($\chi^2_{\text{red}}$) or the corrected Akaike information criterion (AICc) was employed [58, 59]. Angular correlations with counts below 20 (which have error distributions which are not well approximated as Gaussian) were fitted with maximum likelihood estimation instead of $\chi^2_{\text{red}}$ minimization.

#### 1. $^{12}$C$(\alpha,\alpha')^{12}$C$^*$ *at* $\theta_{lab} = 0^\circ$ *and* $E_\alpha = 200$ *MeV.*

For the $^{12}$C$(\alpha,\alpha')^{12}$C measurement at $\theta_{\text{lab}} = 0^\circ$, the population of the Hoyle state at $E_x = 7.654$ MeV yields an approximate kinetic-energy range for the $^{12}$C recoil nucleus of 30 to 110 keV. In comparison to $^{14}$C$(p,t)^{12}$C$^*$, the recoil nucleus for the $^{12}$C$(\alpha,\alpha')^{12}$C$^*$ reaction possess relatively low momentum. Consequently, for the $^{12}$C$(\alpha,\alpha')^{12}$C$^*$ measurement, $\alpha_0$ decay cannot be measured below $E_x \approx 10$ MeV from in the laboratory inertial reference frame due to the CAKE electronic threshold of 600 keV. Neither the inclusive nor the charged-particle decay gated focal-plane spectra from this measurement were included in the **R**-matrix analysis as the primary peak of the Hoyle state was not fully positioned within the full-acceptance range of the spectrometer.

In order to determine the angular correlations of $\alpha_0$ decay, a simulation of the CAKE was implemented with the GEANT4 simulation toolkit [60–62]. To model the populating reaction, the CHUCK3 coupled-channels nuclear reactions code was used and the results were incorporated into the ANGCOR angular-correlation program [63] to determine the $m$-state population amplitudes of populated excited states in $^{12}$C. This method to produce predicted angular correlations of decay was also employed by an analogous study of charged-particle decay from well-resolved $^{16}$O states, which employed an identical experimental setup [56].

The angular correlations of $\alpha_0$ decay from the excitation-energy ranges of $E_x = 10.0$ to 10.3 MeV and $E_x = 10.6$ to 11.1 MeV are respectively presented in panels (a) and (b) in Fig. 9 for the $^{12}$C$(\alpha,\alpha')^{12}$C measurement at $\theta_{\text{lab}} = 0^\circ$. The observed loci corresponding to $\alpha_0$ decay from contaminant $^{16}$O states yield negligible contributions to the analyzed excitation-energy ranges.

For $E_x = 10.0$ to 10.3 MeV, it is observed that the data are not well reproduced with pure $\ell = 0$ or $\ell = 2$ $\alpha_0$ decay. However, an incoherent sum of $\ell = 0$ and $\ell = 2$ decay yields a good fit with $\chi^2_{\text{red}} = 0.959$. This result independently reaffirms the existence of the broad $2^+_2$ state and is consistent with the studies of analogous $^{12}$C$(\alpha,\alpha')^{12}$C measurements in Refs. [25, 26], which indicated that the primary contributions to the excitation-energy range of $E_x = 10.0$ to 10.3 MeV are a broad monopole strength and the $2^+_2$ state located at $E_x = 9.870(60)$ MeV. The excitation-energy range for this angular correlation was chosen to mitigate effects from the electronic threshold of the CAKE (lower limit) and contamination from the $1^-_1$ state located at $E_x = 10.847(4)$ MeV.

For $E_x = 10.6$ to 11.1 MeV, it is observed that the data are not well reproduced with pure $\ell = 0$ $\alpha_0$ decay. A substantially better fit is obtained with an incoherent sum of predicted $\ell = 0$ and $\ell = 1$ $\alpha_0$ distributions, yielding $\chi^2_{\text{red}} = 1.73$. An improved reproduction of the data was obtained with a coherent sum of predicted $\ell = 0$, $\ell = 1$ and $\ell = 2$ $\alpha_0$ distributions, where the coefficients were restricted to be real and yielded $\chi^2_{\text{red}} = 0.870$. This is consistent with the current understanding that the primary contributions to the excitation-energy range of $E_x = 10.6$ to 11.1 MeV should be a broad monopole strength, the high-energy tail of the $2^+_2$ state and the $1^-_1$ state located at $E_x = 10.847(4)$ MeV.

In summary, charged-particle decay from the measurement of $^{12}$C$(\alpha,\alpha')^{12}$C$^*$ at $\theta_{\text{lab}} = 0^\circ$ ($E_\alpha = 200$) enabled the identification of the $2^+_1$ and $1^-_1$ states in addition to broad monopole strength within the broad unresolved structures at $E_x \approx 10$ and 11 MeV, respectively. A consistent model for $^{12}$C must therefore yield non-negligible strengths for these states from the analogous measurements of $^{12}$C$(\alpha,\alpha')^{12}$C at $\theta_{\text{lab}} = 0^\circ$ with $E_\alpha = 118$ and 160 MeV.

#### 2. $^{14}$C$(p,t)^{12}$C$^*$ *at* $\theta_{lab} = 0^\circ$ *and* $E_{beam} = 100$ *MeV.*

The population of the Hoyle state in $^{12}$C through the $^{14}$C$(p,t)^{12}$C measurement at $\theta_{\text{lab}} = 0^\circ$ yields $^{12}$C recoil nuclei with a kinetic energy of $\approx 2.83$ MeV and a polar-angle range of $\theta_{\text{lab}} \gtrapprox 175.5^\circ$. This substantial boost in the laboratory frame, combined with the available decay energy, enabled $\alpha_0$ decays from $^{12}$C to be measured below $E_x \approx 9.0$ MeV with the CAKE, which exhibited an electronic threshold of 750 keV. Since the ion optics of the K600 spectrometer were configured to vertically focus ejectiles, the momentum vector of the recoil nucleus cannot be reconstructed, meaning that the center-of-mass angular correlations could also not be reconstructed. Instead, the angular correlations across the detector channels of the CAKE were analyzed in the laboratory frame. The detection efficiencies of charged-particle decays were simulated assuming particular angular correlations for the decay in the recoil center-of-mass system. Charged-particle gated focal-plane spectra from this measurement were not included in the global fitting analysis due to the aforementioned complications with determining detection efficiencies as well as the limited acceptance of the $\alpha_0$ locus for the primary peak of the Hoyle state (due to electronic thresholds of the CAKE).

For states with $J > 0$, the angular correlations of decay in the center-of-mass of the $^{12}$C recoil nucleus were determined only with $m = 0$ magnetic substates; a valid approximation since the detected scattered particle was detected near $\theta_{\rm lab} = 0°$. To predict the angular correlations of decay in the laboratory inertial reference frame, the angular correlation of excited $^{12}$C nuclei were determined using an assumed angular correlation for the ejectile within the polar-angle range of $\theta_{\rm lab} < 2°$. A GEANT4 simulation of the experimental setup and kinematics were used to predict the angular correlations of decay observed with the CAKE.

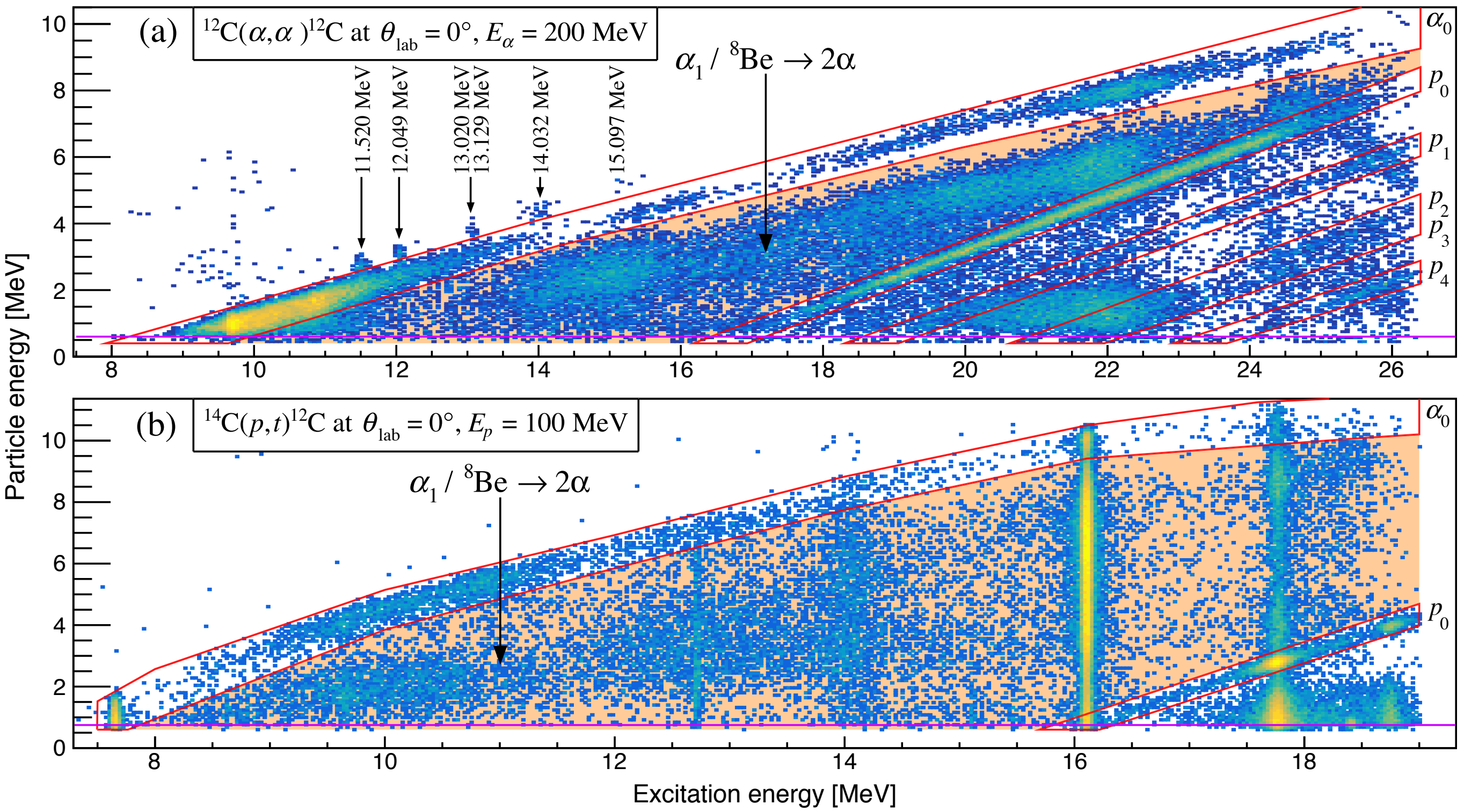

FIG. 8. (Color online) The matrices of particle energy (measured with the CAKE) versus the excitation energy of $^{12}$C for the measurements of (a) $^{12}$C$(\alpha,\alpha')^{12}$C at $\theta_{\rm lab} = 0°$ ($E_\alpha = 200$ MeV) and (b) $^{14}$C$(p,t)^{12}$C at $\theta_{\rm lab} = 0°$. The matrices in panels (a) and (b) are gated on a subset of the CAKE detector channels, corresponding to decay-angle ranges of $140° < \theta_{\rm lab} < 161°$ and $153° < \theta_{\rm lab} < 165°$, respectively. The red lines indicate the observed $\alpha_0$ and proton decay modes of $^{12}$C. The orange-highlighted regions indicate $\alpha$ particles corresponding to either $\alpha_1$ decay or the $^8$Be $\rightarrow 2\alpha$ breakup channel. The violet, horizontal lines in panels (a) and (b) indicate the approximate electronic thresholds for the silicon detectors of 600 and 750 keV, respectively. For panel (a), loci corresponding to $\alpha_0$ decay from contaminant states of $^{16}$O are labeled and a display in color threshold of $> 2$ events is imposed.

The angular correlation of $\alpha_0$ decay for various excitation-energy ranges are presented in Fig. 10. The data were fitted with a maximum likelihood estimation for decay with orbital angular momenta of $\ell = 0$, $\ell = 2$ as well as an incoherent sum of the two distributions. For $E_x = 8.5$ to 9.0 MeV, shown on panel (a) of Fig. 10, data points at polar angles below $\theta_{\rm lab} = 145°$ were omitted as they were affected by the electronic threshold of the CAKE. The AICc estimators indicate that the best quality model corresponds to pure $\ell = 0$ decay and that the $\ell = 2$ contribution is negligible. This is in agreement with the current understanding that the excitation-energy region of $E_x \approx 8.5$ to 9 MeV is dominated by monopole strength.

The angular correlation of $\alpha_0$ decay from the $J^\pi = 3^-$ state located at $E_x = 9.641(5)$ MeV is presented in panel (b) of Fig. 10. Whilst this relatively narrow state is well resolved from the surrounding broad $J^\pi = 0^+$ and $2^+$ resonances, its relatively low population required data at multiple angles to be combined for reliable peak fitting. The best-fitting $\ell = 3$ contribution is consistent with the $J^\pi = 3^-$ nature of the state, however the reduced granularity yields a poor discrimination from the $\ell = 0$ contribution.

The angular correlation of $\alpha_0$ decay from $E_x = 9.8$ to 10.0 MeV is presented in panel (c) of Fig. 10. It is observed that the data are not well reproduced with either pure $\ell = 0$ or $\ell = 2$ $\alpha_0$ decay. The AICc estimators indicate that the best quality model corresponds to the incoherent sum of $\ell = 0$ and $\ell = 2$ decay and the two components are of similar strength. This is consistent with the current understanding that this excitation-energy region should contain broad monopole strength as well as the $2_2^+$ state which has a listed resonance energy of 9.870(60) MeV [25, 26]. The excitation-energy range for this angular correlation was chosen to mitigate

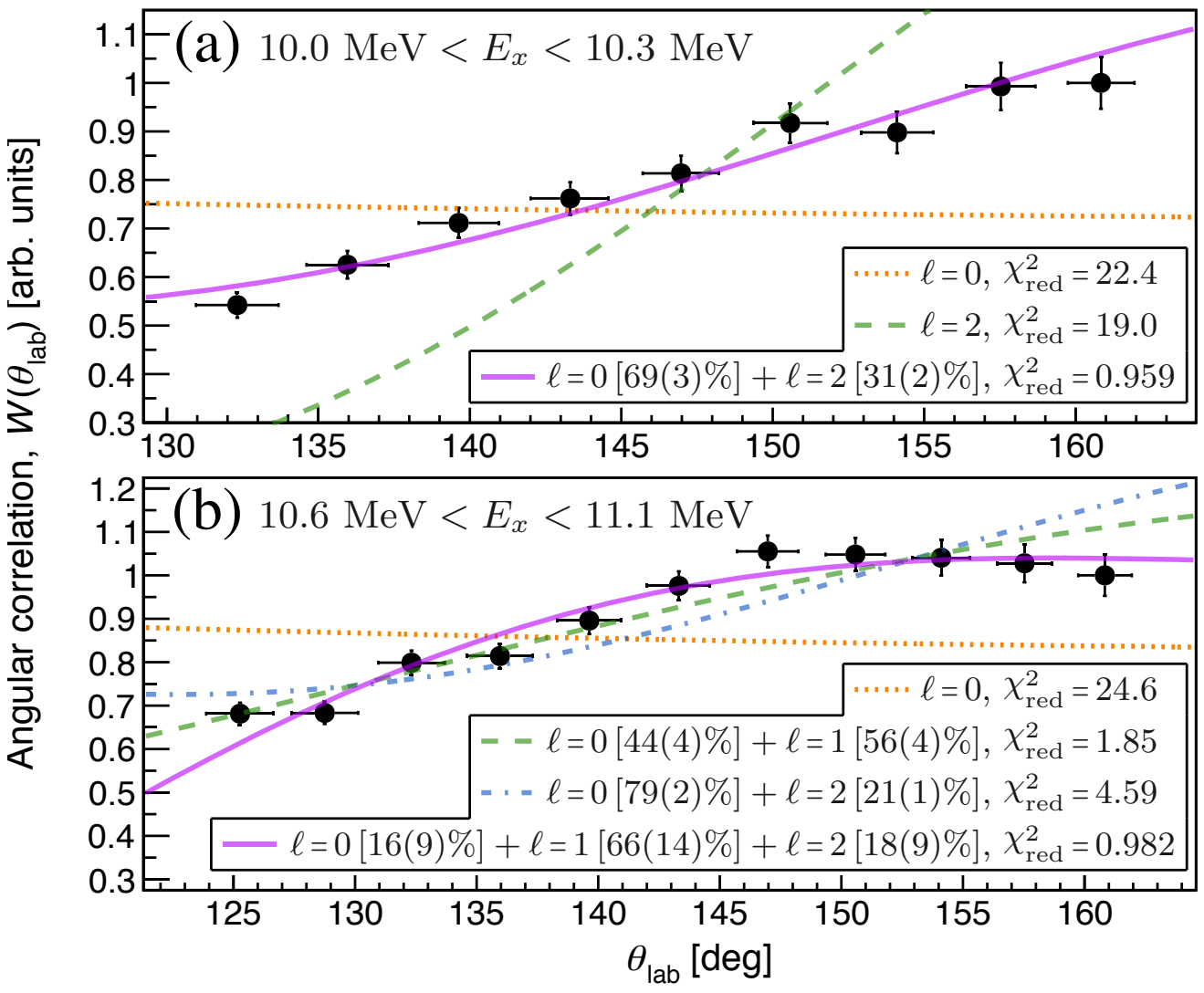

FIG. 9. Angular-correlation functions of $\alpha_0$ decay across the ring detector channels of the CAKE, relative to the beam axis, for the excitation-energy regions of (a) $E_x = 10.0$ to 10.3 MeV and (b) $E_x = 10.6$ to 11.1 MeV from the measurement of $^{12}$C$(\alpha,\alpha')^{12}$C at $\theta_{\rm lab} = 0^\circ$ with $E_\alpha = 200$ MeV. The sums of angular correlations with different $\ell$-values are incoherent, with the exception being the combination of $\ell = 0$, 1 and 2 in panel (b).

contamination from the $3_1^-$ and $1_1^-$ states.

The angular correlation of $\alpha_0$ decay from $E_x = 10.6$ to 11.1 MeV is presented in panel (d) of Fig. 10. It is observed that the data is best described by a coherent sum of $\ell = 0, 1$ and 2 contributions. This is consistent with the understanding that the $J^\pi = 1_1^-$ state at $E_x = 10.847(4)$ MeV should contribute in addition to the underlying broad monopole and $2_2^+$ strength in this excitation-energy region.

The angular correlation of $\alpha_0$ decay from the $J^\pi = 2^+$ state located at $E_x = 16.106$ MeV is presented in panel (e) of Fig. 10. Data points at polar angles below $\theta_{\rm lab} = 149^\circ$ were omitted as they corresponded to $\alpha_0$ loci which could not be well-resolved from the broad $\alpha_1/^8$Be $\rightarrow 2\alpha$ locus, as shown in panel (b) of Fig. 8. The clear $\ell = 2$ identification of the $\alpha_0$ decay from this strongly populated, well-resolved state verifies the method of $\ell$-value decomposition employed for this analysis.

The angular correlation of $\alpha_0$ decay from the $J^\pi = 0^+$ state located at $E_x = 17.760$ MeV is presented in panel (f) of Fig. 10. Data points at polar angles above $\theta_{\rm lab} = 140^\circ$ were omitted as they corresponded to $\alpha_0$ loci which were affected by saturation of the electronic signals from the CAKE. The clear $\ell = 0$ identification of the $\alpha_0$ decay from this strongly populated, well-resolved state verifies the method of $\ell$-value decomposition employed for this analysis.

In summary, charged-particle decay from the measurement of $^{14}$C$(p,t)^{12}$C$^*$ at $\theta_{\rm lab} = 0^\circ$ enabled the study of $\alpha_0$ decay in the excitation-energy range of $E_x = 8.5$ to 9.0 MeV - a region which cannot be accessed with the measurement of $^{12}$C$(\alpha,\alpha')^{12}$C$^*$ at $\theta_{\rm lab} = 0^\circ$ ($E_\alpha = 200$ MeV) due to the electronic thresholds of the CAKE. The angular correlation indicates that the observed broad strength is purely monopole and there is no statistically significant evidence to introduce non-monopole strength in this excitation-energy region. At $E_x \approx 10.0$ MeV, the data indicate that both monopole and $2_2^+$ states contribute with comparable strengths - a result which must be corroborated by the fit of a consistent model for $^{12}$C.

## IV. MODELS AND RESULTS

Various models were investigated to describe the broad structures in the excitation-energy region of $E_x = 7$ to 13 MeV. These models assumed different sources of monopole strength above the Hoyle state and different permutations of interference. All the inclusive spectra for the measurements listed in Table I are included in the global fit, with the exception being $^{12}$C$(\alpha,\alpha')^{12}$C measurement at $\theta_{\rm lab} = 0^\circ$ with $E_\alpha = 200$ MeV as the primary peak of the Hoyle state was not fully accepted on the focal plane. For each fitted excitation-energy region, all relevant $^{12}$C states listed in the ENSDF database were included [1]. In particular, the corresponding fits all included the well-established $J^\pi = 2_2^+$ and $3_1^-$ states, located at $E_x = 9.870(60)$ and 9.641(5) MeV, respectively. The observed resonance energies and widths were typically restricted within $3\sigma$ of the listed values. However, in certain cases this was extended to $5\sigma$ as some previous studies employed simplistic analyses, which may have introduced significant systematic error. For the fits, the total width of the Hoyle state was fixed to $\Gamma = 9.3$ eV at $E_x = 7.65407$ MeV (unless explicitly stated). For the $2_2^+$ state, a recent **R**-matrix analysis of photodisintegration data produced $E_r = 10.025(50)$ MeV and $\Gamma = 1.60(13)$ MeV [9, 27, 64], which differ substantially from the values of $E_x = 9.870(60)$ and $\Gamma = 850(85)$ keV listed in the ENSDF database [1]. Since photodisintegration does not populate the surrounding broad monopole strength which often complicates the analysis of the $2_2^+$ state, it is deduced that results of Refs. [9, 27, 64] may be more reliable. Consequently, a conservatively large range is permitted for the total width of the $2_2^+$ state: $3\sigma$ below $\Gamma = 850(85)$ keV and $3\sigma$ above $\Gamma = 1.60(13)$ MeV.

To fit the inclusive spectra, a maximum likelihood estimation was employed instead of a $\chi^2_{\rm red}$ minimization as there are excitation-energy regions with low counts where the associated errors cannot be well approximated as Gaussian. To determine the best quality model to describe the inclusive data, the Akaike information criterion (AIC) and Bayesian information criterion (BIC) were employed [58, 65]. These estimators account for the goodness of fit to the data whilst penalizing the number of estimated parameters, thus enabling the model with the lowest AIC/BIC to be selected as the best quality

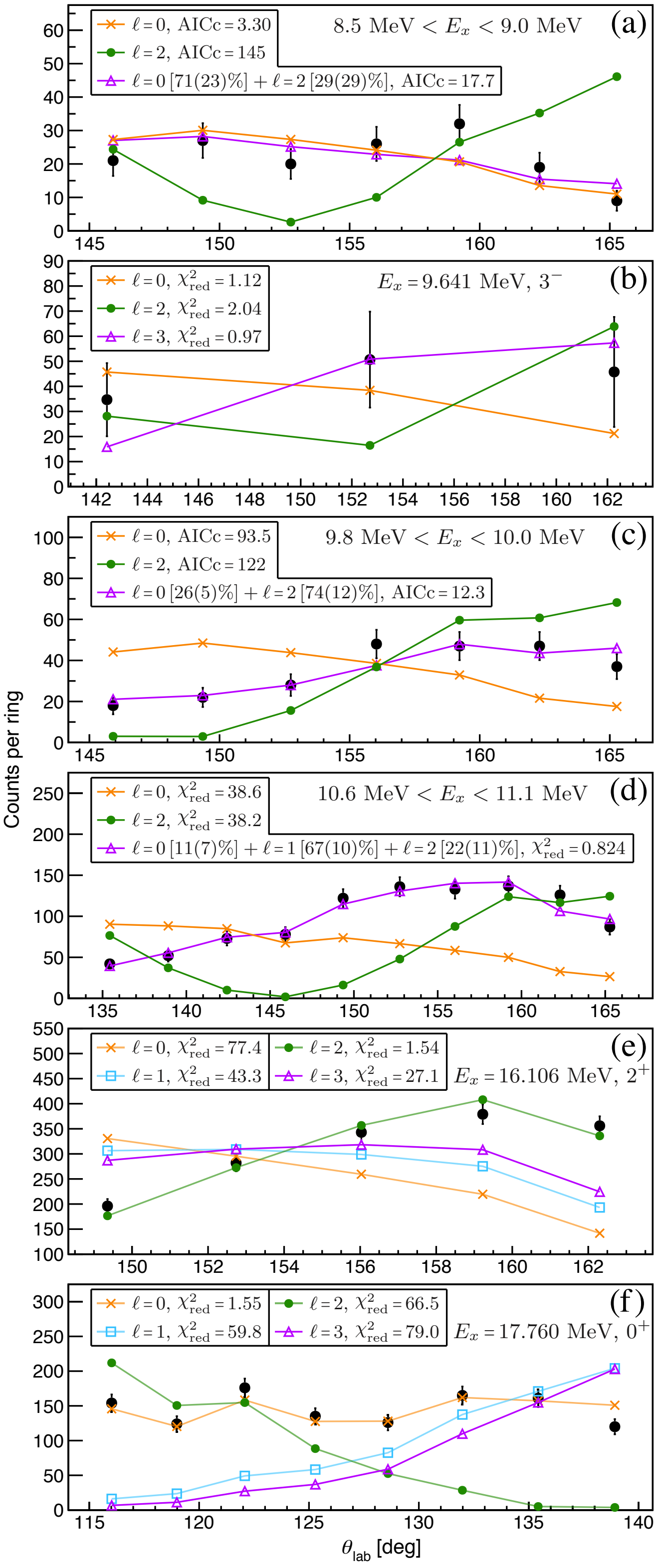

FIG. 10. The angular correlations of $\alpha_0$ decay across the ring detector channels of the CAKE, relative to the beam axis, for the excitation-energy regions of (a) $E_x = 8.5$ to 9.0 MeV and (b) $E_x = 9.8$ to 10.0 MeV as well as (c) the narrow $2^+$ state at $E_x = 16.106$ MeV from the measurement of $^{14}\mathrm{C}(p,t)^{12}\mathrm{C}$ at $\theta_{\rm lab} = 0°$ with $E_{\rm beam} = 100$ MeV. The sums of angular correlations with different $\ell$-values are incoherent, with the exception being the combination of $\ell = 0$, 1 and 2 in panel (d).

model.

As this work is focused on unraveling the sources of monopole strength in the excitation-energy region of $E_x = 7$ to 13 MeV, the investigated models are labeled by the monopole resonances which are considered. For example, model $M_{0_2^+}$ considers only the $0_2^+$ Hoyle state in this excitation-energy region, whilst $M_{0_2^+0_3^+}$ considers both the $0_2^+$ Hoyle state and previously established $0_3^+$ state. If interference is considered to occur between monopole states, this is indicated by linking dashed green lines, e.g. $M_{0_2^+0_3^+}$.

In this section, the presented results correspond to analyses which employ the penetrability prescription of Equation 5. For completeness, an analogous set of results using the penetrability prescription of Equation 6 is also presented in the appendix as this alternative prescription has been used in other studies. It was found that the two prescriptions of 5 and 6 yield similar fit results which lead to the same conclusions.

### A. Model $M_{0_2^+}$: Hoyle state only

In this model, it is assumed that the broad monopole strength that is consistently observed above the primary peak of the $0_2^+$ Hoyle state and below $E_x \approx 13$ MeV corresponds only to the ghost of the Hoyle state. The optimized fit with this model is presented in Fig. 11 with the results summarized in Table IV.

It is observed that this model yields a poor reproduction of the data with particularly large differences at $E_x \approx 9$ and $\approx 11.5$ MeV for the $^{12}\mathrm{C}(\alpha,\alpha')^{12}\mathrm{C}$ measurements at $\theta_{\rm lab} = 0°$. The fitted polynomial backgrounds in panels (a) and (b) of Fig. 11 do not appear reasonable as they are simultaneously fitted on the relevant instrumental background spectra in Fig. 7, which restricts their shape. Another problem with this model is the considerably larger yield for the $2_2^+$ state with respect to the monopole strength at $E_x \approx 10$ MeV for the $^{14}\mathrm{C}(p,t)^{12}\mathrm{C}$ reaction at $\theta_{\rm lab} = 21°$. For this excitation-energy region, the angular correlation of $\alpha_0$ decay, presented in panel (b) of Fig. 10, indicates that the contribution from the broad monopole strength should be similar to that of the $2_2^+$ state. The fit produces an optimized $\alpha_1$ width which is vanishingly small and consequently, the lineshape of the Hoyle state reduces to that of a single-level, single-channel approximation - a simple description that is often used for the Hoyle state and its associated ghost [13]. Since the penetrability is a monotonic function, this (functionally) single-channel description cannot reproduce the double-peaked broad monopole strength above the Hoyle state that is observed in the MDA by Itoh *et al.* [25].

### B. Model $M_{0_2^+0_3^+}$:

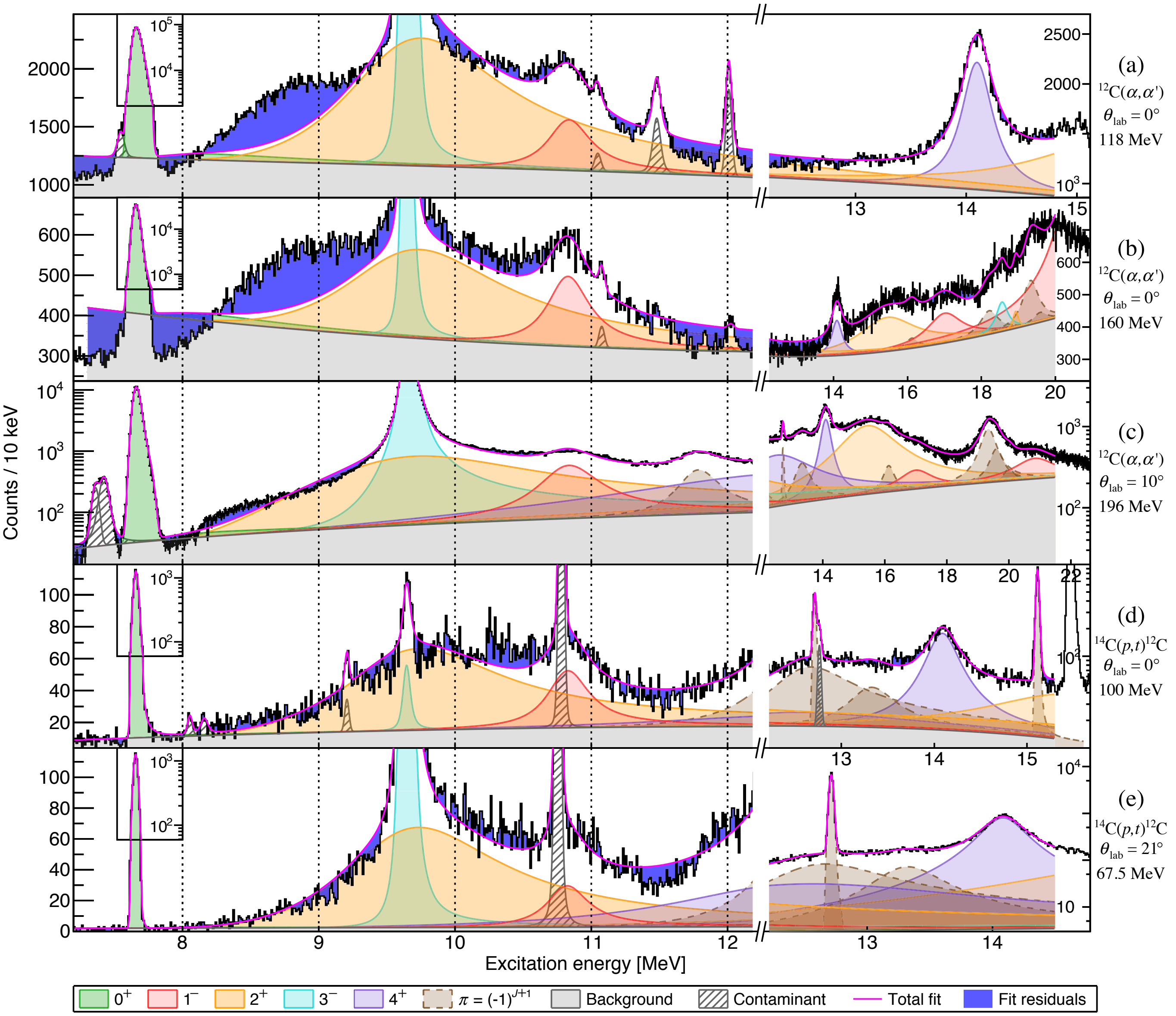

FIG. 11. The optimized global fit for model $M_{0_2^+}$: it is assumed that the broad monopole strength above the primary peak of the $0_2^+$ Hoyle state and below $E_x \approx 13$ MeV corresponds only to the ghost of the Hoyle state. See text for details.

**the Hoyle state and a broad $0_3^+$ state at $E_x \approx 10$ MeV**

Two distinct $0^+$ resonances are assumed to contribute to the excitation-energy region of $E_x = 7$ to 13 MeV: the Hoyle state and a broad resonance previously observed at $E_x \approx 10$ MeV with a width of $\Gamma \approx 3$ MeV. The ENSDF database lists two $0^+$ resonances situated at $E_x = 9.930(30)$ and 10.3(3) MeV, with widths of $\Gamma = 2.710(80)$ and 3.0(7) MeV, respectively [1]. The latter resonance corresponds to measurements of $\beta$-decay from $^{12}$B and $^{12}$N [2], and of the $^{12}$C$(\alpha, \alpha')^{12}$C reaction with $E_{\rm beam} = 120$ MeV at $0° < \theta_{\alpha'} < 4.5°$ [3]. The close proximity of these two resonance energies, with respect to their relatively large widths, suggests that these two listed resonances are one and the same [66]. This broader resonance should exhibit considerable overlap with the ghost of the Hoyle state and model $M_{0_2^+ 0_3^+}$ permits interference between these two resonances according to the two-level approximation shown in Equation 10. The interference terms of the monopole lineshape depend on the relative signs for reduced-width amplitudes of the $\alpha_0$ and $\alpha_1$ exit channels: $\gamma_{1,\alpha_0}$, $\gamma_{1,\alpha_1}$, $\gamma_{2,\alpha_0}$ and $\gamma_{2,\alpha_1}$, which are denoted as:

$$\begin{array}{c} \\ \alpha_0 \\ \alpha_1 \end{array} \begin{array}{c} \begin{array}{cc} 0_2^+ & 0_3^+ \end{array} \\ \left[ \begin{array}{cc} + & - \\ + & - \end{array} \right] \end{array},$$

respectively. Equation 2 indicates that for a particular exit channel, inverting the signs of the reduced-width amplitudes produces an identical spectral lineshape, e.g. $\left[\begin{smallmatrix}+&-\\-&+\end{smallmatrix}\right] \equiv \left[\begin{smallmatrix}-&+\\+&-\end{smallmatrix}\right]$, and similarly for three (or more) interfering resonances. As mentioned in Section III A, for two interfering resonances, the signs of the reduced-width amplitudes for a particular decay channel are directly related to the form of interference in the excitation-energy region between the respective resonance energies. To be more physically descriptive, the different permutations of interference for model $M_{0_2^+0_3^+}$ are also equivalently labeled with the form of interference for each exit channel (between the resonance energies). For example, the case of interfering $0_2^+$ and $0_3^+$ states with constructive $\alpha_0$ and destructive $\alpha_1$ interference is denoted as $M_{0_2^+0_3^+(+\alpha_0,-\alpha_1)} \equiv M_{0_2^+0_3^+}\left[\begin{smallmatrix}+&-\\-&-\end{smallmatrix}\right]$.

TABLE IV. Summary of the optimal fit results with the penetrability prescription of Equation 5. Errors for the resonance energies account for both the fit error and an estimated focal-plane calibration error whilst the errors for the widths are purely from the fit.

| Model | AIC<br>BIC | State | $E_r$<br>[MeV] | $\Gamma_{\alpha_0}(E_r)$<br>[keV] | $\theta^2_{\alpha_0}$ | $r_{\alpha_0}$<br>[fm] | $\Gamma_{\alpha_1}(E_r)$<br>[keV] | $\theta^2_{\alpha_1}$ | $r_{\alpha_1}$<br>[fm] | $\Gamma(E_r)$<br>[keV] | $\Gamma_{\text{FWHM}}$<br>[keV] |
|---|---|---|---|---|---|---|---|---|---|---|---|
| $M_{0_2^+}$ | 29820 | $0_2^+$ | [a]7.65407 | [a]$9.3\times10^{-3}$ | 0.052 | 11 | $\approx 0$ | $\approx 0$ | 11 | $9.3\times10^{-3}$ | |
| | 29470 | $2_2^+$ | 9.869(8) | 1661(13) | 1.057(8) | 11 | $\approx 0$ | $\approx 0$ | 11 | 1661(13) | 1451(9) |
| $M_{0_2^+0_3^+}$ | 15833 | $0_2^+$ | [a]7.65407 | [a]$9.3\times10^{-3}$ | 0.14 | 9 | $\approx 0$ | $\approx 0$ | 9 | $9.3\times10^{-3}$ | |
| $\left[\begin{smallmatrix}+&+\\-&-\end{smallmatrix}\right]$ | 15467 | $2_2^+$ | 10.098(44) | 1748(126) | 0.964(69) | 9 | $\approx 0$ | $\approx 0$ | 9 | 1748(126) | 1465(75) |
| | | $0_3^+$ | 10.197(8) | $\approx 2279$ | $\approx 1$ | 9 | $\approx 0$ | $\approx 0$ | 9 | $\approx 2279$ | |
| $M_{0_2^+0_\Delta^+0_3^+}$ | 12827 | $0_2^+$ | [a]7.65407 | [a]$9.3\times10^{-3}$ | 0.052 | 11 | $\approx 0$ | $\approx 0$ | 11 | $9.3\times10^{-3}$ | |
| | 12445 | $0_\Delta^+$ | 9.553(34) | 3375(167) | 2.110(104) | 11 | $\approx 0$ | $\approx 0$ | 11 | 3375(167) | |
| $\left[\begin{smallmatrix}-&+&-\\-&-&+\end{smallmatrix}\right]$ | | $2_2^+$ | 9.893(26) | 1026(85) | 0.647(53) | 11 | 416(54) | 5.15(67) | 11 | 1443(101) | 960(45) |
| | | $0_3^+$ | 10.921(15) | 1454(197) | 0.617(84) | 11 | 679(125) | 3.22(59) | 11 | 2133(233) | |
| $M_{0_2^+0_\Delta^+0_3^+}$ | 12828 | $0_2^+$ | [a]7.65407 | [a]$9.3\times10^{-3}$ | 0.24 | 08 | $\approx 0$ | $\approx 0$ | 08 | $9.3\times10^{-3}$ | |
| | 12446 | $0_\Delta^+$ | 9.439(30) | 4624(58) | 3.04(4) | 11 | 54(13) | 1.7(4) | 11 | 4678(59) | 2179(16) |
| $\left[\begin{smallmatrix}+&+\\-&-\end{smallmatrix}\right]$ | | $2_2^+$ | 9.913(18) | 1553(67) | 0.951(41) | 08 | 146(1) | 2.72(2) | 08 | 1699(67) | 1057(26) |
| | | $0_3^+$ | 10.990(10) | 1102(47) | 0.363(15) | 08 | $\approx 0$ | $\approx 0$ | 08 | 1102(47) | |

[a] Fixed parameters in the fit optimisation.

Fig. 12 presents a comparison of the fits for the measurements of $^{12}$C$(\alpha,\alpha')^{12}$C at $\theta_{\text{lab}} = 0°$, which are the most selective for monopole strengths. It is observed that model $M_{0_2^+0_3^+}$ provides a significantly better fit than $M_{0_2^+}$. However, there is still a significant underestimation of the data at $E_x \approx 9$ (see Fig. 12). The best-quality submodels which account for both the $0_2^+$ and $0_3^+$ states correspond to the fits with submodels $M_{0_2^+0_3^+(+\alpha_0,+\alpha_1)}$ and $M_{0_2^+0_3^+(+\alpha_0,-\alpha_1)}$, which are in contrast to submodels $M_{0_2^+0_3^+(-\alpha_0,+\alpha_1)}$ and $M_{0_2^+0_3^+(-\alpha_0,-\alpha_1)}$ which both provide similarly poor fits. Since the $\alpha_1$ decay channel is substantially smaller than that of the $\alpha_0$ decay channel between $E_x = 7$ and 11 MeV (even when accounting for the broad width of the $2_1^+$ $^8$Be daughter state), it is observed that the interference through the $\alpha_0$ decay channel plays the dominant role in providing a good fit. Submodel $M_{0_2^+0_3^+(+\alpha_0,+\alpha_1)}$ corresponds to the best-fitting permutation of $\alpha_0$ and $\alpha_1$ interference, with the results summarized in Table IV with the decomposition given in Fig. 13. Whilst model $M_{0_2^+0_3^+(+\alpha_0,+\alpha_1)}$ provides a substantially better fit, a clear systematic underestimation of the data is still observed at $E_x \approx 9$ MeV for the measurements of $^{12}$C$(\alpha,\alpha')^{12}$C at $\theta_{\text{lab}} = 0°$. Notably, no such discrepancy at $E_x \approx 9$ MeV was observed for the fit of $^{14}$C$(p,t)^{12}$C at $\theta_{\text{lab}} = 0°$. In this particular fit, the $\alpha_0$ width of the $0_3^+$ state is constrained by the Wigner limit. Allowing all broad states between $E_x = 7$ and 13 MeV to exceed the Wigner limits yields a slightly improved fit presented in Fig. 14. However, a clear deficit in the predicted yield at $E_x \approx 9$ MeV persists for the measurements with $^{12}$C$(\alpha,\alpha')^{12}$C at $\theta_{\text{lab}} = 0°$. This indicates that the excess monopole strength at $E_x \approx 9$ MeV corresponds to a highly collective excitation as it is more selectively populated with inelastic alpha scattering in comparison to neutron-pair transfer, which should not access the 1p-1h excitations that constitute a breathing-mode excitation.

To test whether a difference in the total width of the Hoyle state may provide an improved fit, the total width of the Hoyle state was tested at $5\sigma$ below and above the listed value of $\Gamma = 9.3(9)$ eV [1] (see Fig. 12). The corresponding $5\sigma$ band in Fig. 12 corresponds to the range spanned by the optimized fits with the Hoyle-state width at $\Gamma = 4.8$ and 13.8 eV. Even at these extreme, statistically unlikely values, there is still clear systematic excess in the data at $E_x \approx 9$ MeV. This region has been shown to be dominated by monopole strength through the angular correlation in panel (a) of Fig. 10 and the MDA of $^{12}$C$(\alpha,\alpha')^{12}$C in Ref. [25]. Another discrepancy of submodel $M_{0_2^+0_3^+(+\alpha_0,+\alpha_1)}$ is shown by the decomposition of fits presented in Figs. 13 and 14, which reveal a highly suppressed strength for the $2_2^+$ state for the measurements with $^{12}$C$(\alpha,\alpha')^{12}$C and $^{14}$C$(p,t)^{12}$C at $\theta_{\mathrm{lab}} = 0°$. This is inconsistent with the MDA of Ref. [25] and the angular correlation of charged-particle decay in panel (a) of Figs. 9 and 10, which reveal that for $^{12}\mathrm{C}(\alpha,\alpha')^{12}\mathrm{C}$ and $^{14}\mathrm{C}(p,t)^{12}\mathrm{C}$ at $\theta_{\mathrm{lab}} = 0°$, the population of the $2_2^+$ state at $E_x \approx 10$ should be similar to that of the broad monopole contribution. Furthermore, the broad monopole lineshapes at $E_x \approx 10$ for $^{12}\mathrm{C}(\alpha,\alpha')^{12}\mathrm{C}$ at $\theta_{\mathrm{lab}} = 0°$ in this work do not qualitatively match the double-peaked monopole strength from the MDA of Ref. [25] (see Figs. 13 and 14). Consequently, model $M_{0_2^+0_3^+}$ is not considered a valid description of the resonances of $^{12}$C between $E_x \approx 7$ to 13 MeV.

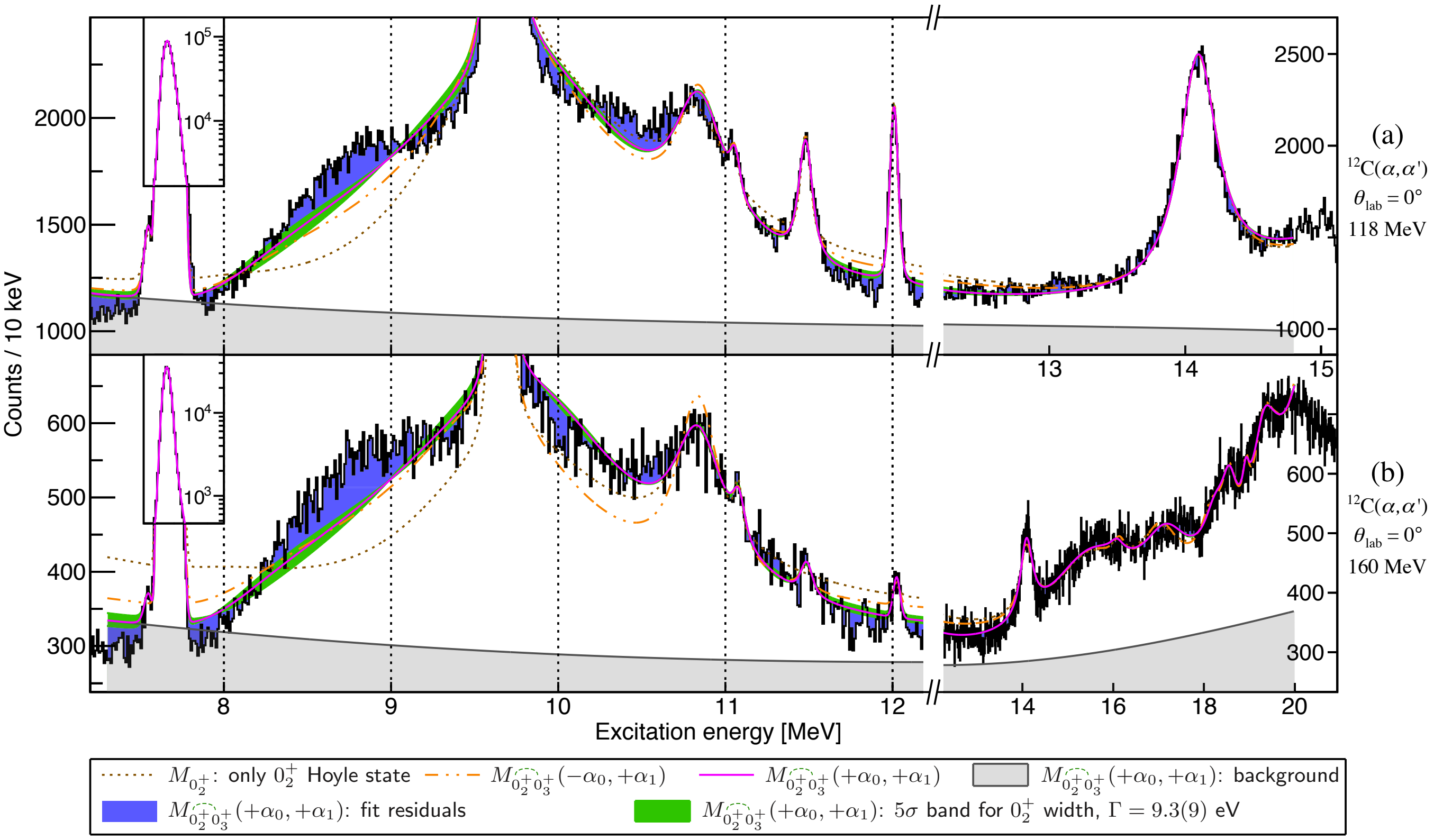

FIG. 12. A comparison of the fits which only include the previously established sources of monopole strength. The 5$\sigma$ band for the $0_2^+$ Hoyle state corresponds to the range between the fits where the total width of the Hoyle state was set at $\pm 5\sigma$ corresponding to $\Gamma = 9.3(9)$ eV. See text for details.

### C. Models $M_{0_2^+0_\Delta^+0_3^+}$ and $M_{0_2^+0_\Delta^+0_3^+}$: the Hoyle state, a broad $0_3^+$ state at $E_x \approx 10$ MeV and additional strength at $E_x \approx 9$ MeV

Three distinct sources of monopole strength contribute to the excitation-energy region of $E_x = 7$ to 13 MeV: the Hoyle state, the broad resonance previously observed at $E_x \approx 10$ MeV, and an additional broad state (denoted $0_\Delta^+$) at $E_x \approx 9$ MeV. The location of this additional source of monopole strength coincides with the deficit in the predicted yield at $E_x \approx 9$ MeV observed with model $M_{0_2^+0_3^+(+\alpha_0,+\alpha_1)}$ for the measurements with $^{12}\mathrm{C}(\alpha,\alpha')^{12}\mathrm{C}$ at $\theta_{\mathrm{lab}} = 0°$. Various theoretical models suggest that this additional $0_\Delta^+$ state may correspond to the breathing-mode excitation of the Hoyle state.

Submodel $M_{0_2^+0_\Delta^+0_3^+}$ permits interference between all three monopole resonances by following the general parametrization of Equation 2. The signs of the reduced-width amplitudes are denoted as

$$\begin{array}{c} \\ \alpha_0 \\ \alpha_1 \end{array} \begin{array}{c} \begin{array}{ccc} 0_2^+ & 0_\Delta^+ & 0_3^+ \end{array} \\ \begin{bmatrix} + & - & - \\ + & + & - \end{bmatrix} \end{array}.$$

In contrast to model $M_{0_2^+0_3^+}$ which only has two monopole states between $E_x \approx 7$ to 13 MeV, the case of three interfering levels does not possess a simple correlation between the signs of the reduced-width amplitudes and the constructive or destructive nature of the interference. Therefore, the permutations of interference are directly labeled with the signs, e.g. $M_{0_2^+0_\Delta^+0_3^+}\left[\begin{smallmatrix} - & + & - \\ - & - & + \end{smallmatrix}\right]$.

If the $0_\Delta^+$ state corresponds to a dynamic, collective monopole response, it is possible that it may not interfere with neighboring $0^+$ states which are understood to be statically deformed. In such a case, submodel $M_{0_2^+0_\Delta^+0_3^+}$ may provide an appropriate parametrization of the monopole strength, whereby only the $0_2^+$ and $0_3^+$ states are permitted to interfere and the additional $0_\Delta^+$ state is parameterized as an isolated state. The permutations of interference for submodel $M_{0_2^+\,0_\Delta^+\,0_3^+}$ are denoted by the nature of interference between the $0_2^+$ and $0_3^+$ states, as described in IV B, e.g. $M_{0_2^+\,0_\Delta^+\,0_3^+}(+\alpha_0,-\alpha_1) \equiv M_{0_2^+\,0_\Delta^+\,0_3^+}\left[\begin{smallmatrix} + & - \\ - & - \end{smallmatrix}\right]$. Since the additional $0_\Delta^+$ state may correspond to the dilute breathing-mode excitation of the Hoyle state, it may require a different (and possibly larger) channel radius with respect to the Hoyle state and its respective rotational-band states. For submodel $M_{0_2^+\,0_\Delta^+\,0_3^+}$, the channel radius for the $0_\Delta^+$ state was therefore tested independently from the global channel-radius parametrization discussed in Section III A 2.

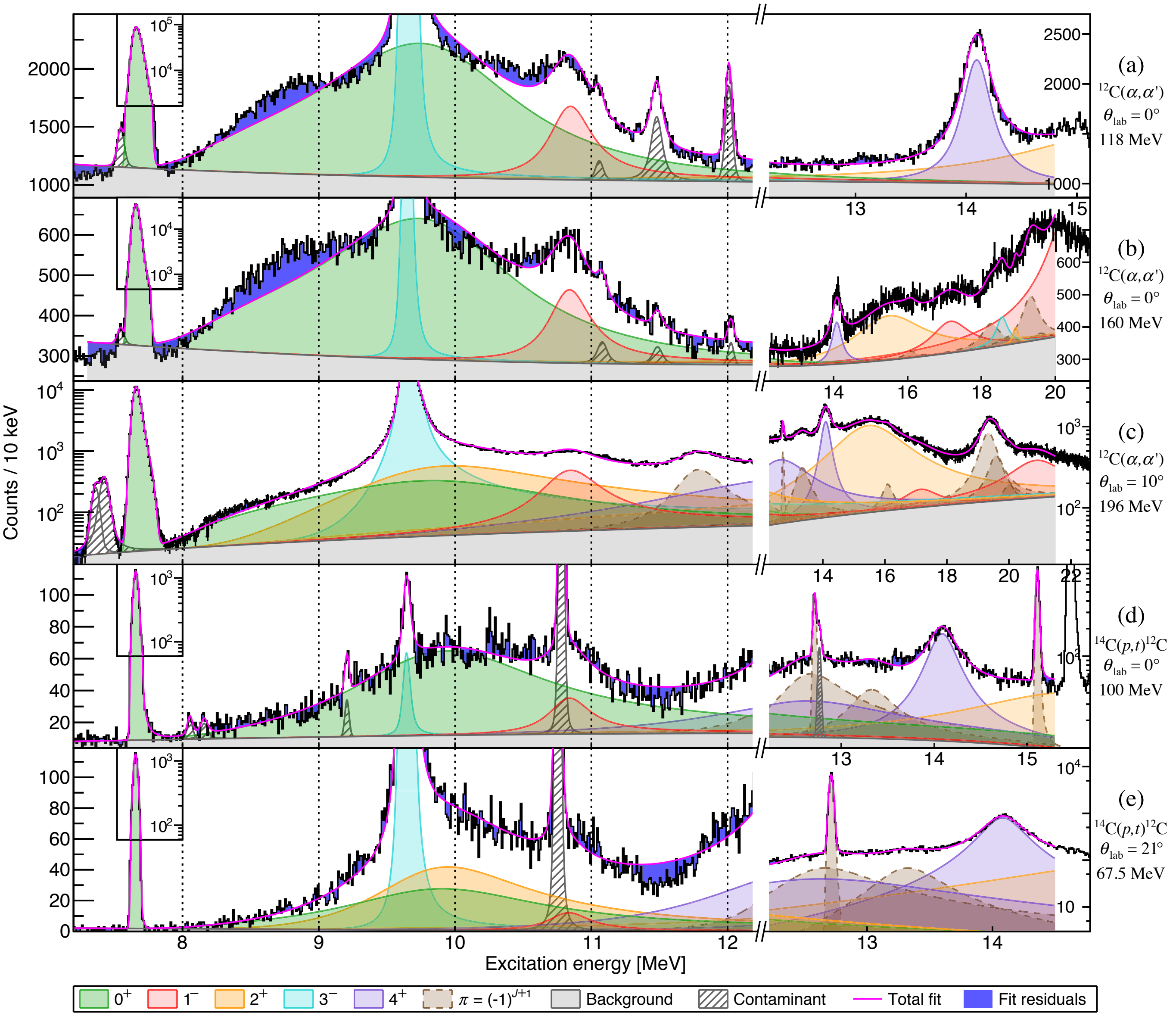

FIG. 13. The optimized global fit for submodel $M_{0_2^+\,0_3^+}(+\alpha_0,+\alpha_1)$, which accounts for all of the previously established sources of monopole strength between $E_x = 7$ and 13 MeV: the $0_2^+$ Hoyle state and a broad $0_3^+$ resonance at $E_x \approx 10$ MeV. This particular fit is used in the comparison in Fig. 12 and the results are summarized in Table IV. See text for details.

The optimal fits for submodels $M_{0_2^+\,0_\Delta^+\,0_3^+}$ and $M_{0_2^+\,0_\Delta^+\,0_3^+}$ are presented in Figs. 15 and 16, respectively, with the results summarized in Table IV. Both submodels yield better-quality fits to the data compared to models which only account for the previously established $0_2^+$ and $0_3^+$ states. Submodel $M_{0_2^+\,0_\Delta^+\,0_3^+}$ yields the best-quality fit to the data, with a good fit at $E_x \approx 9$ MeV for $^{12}$C$(\alpha,\alpha')^{12}$C at $\theta_{\rm lab} = 0°$, as shown in Fig. 15. This is in contrast to submodel $M_{0_2^+\,0_3^+}(+\alpha_0,+\alpha_1)$:which produced a clear deficit in the predicted yield at $E_x \approx 9$ MeV. For submodels $M_{0_2^+\,0_\Delta^+\,0_3^+}$ and $M_{0_2^+\,0_\Delta^+\,0_3^+}$, the many permutations for the signs of the reduced width amplitudes, combined with the different tested channel radii, yield several relatively similar fits to the inclusive spectra. However, whilst

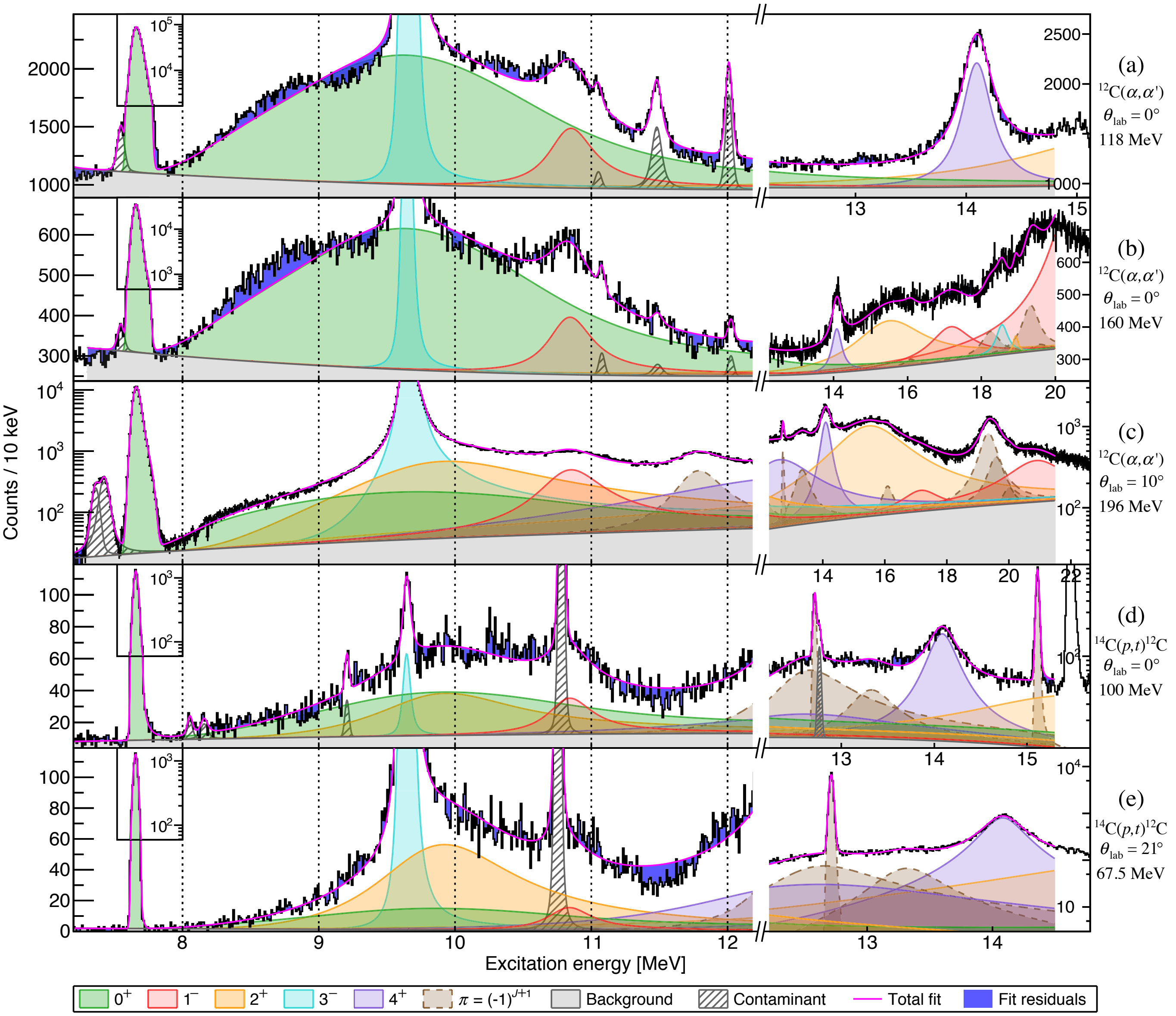

FIG. 14. The optimized global fit for submodel $M_{0_2^+ 0_3^+(+\alpha_0,+\alpha_1)}$, which accounts for all of the previously established sources of monopole strength between $E_x = 7$ and 13 MeV: the $0_2^+$ Hoyle state and a broad $0_3^+$ resonance at $E_x \approx 10$ MeV. In this particular case, the Wigner limit is not applied to any broad states between $E_x = 7$ to 13 MeV. See text for details.

the inclusive fits are similar, the various broad contributions between $E_x \approx 7$ to 13 MeV differ substantially in their shapes, positions and relative strengths. As discussed in Sections III D 1 and III D 2, since both the $\alpha_0$ and $\alpha_1$ charged-particle gated spectra are not simultaneously fitted in this work, it is not possible to place strict constraints on which fits are valid. To account for this systematic fitting error for submodels $M_{0_2^+ 0_\Delta^+ 0_3^+}$ and $M_{0_2^+ 0_\Delta^+ 0_3^+}$, a set of recommended observable values were determined by averaging over fits with different signs for the reduced-width amplitudes and various channel radii. For the fits in this work, it was found that the parameters for the broad states between $E_x = 7$ and 13 MeV are heavily correlated and consequently, the fit errors for the resonance energies and (reduced) widths alone do not capture the full extent of the fitting errors. It is understood that the systematic fitting errors discussed above provide an improved assessment of the total analysis errors in this work. The following conditions were used to determine which fits qualify to be averaged:

**AIC estimator:** for the recommended observable parameters of a particular submodel, the fits which qualify to be averaged must possess AIC estimators which are within 1% of the optimal AIC estimator for the submodel. This ensures that only relatively high quality models for the inclusive spec-tra are considered. The 1% range corresponds to fluctuations in the inclusive fits which are on the order of the estimated systematic experimental errors/artefacts for the focal-plane spectra (e.g. small inconsistencies in VDC efficiency).

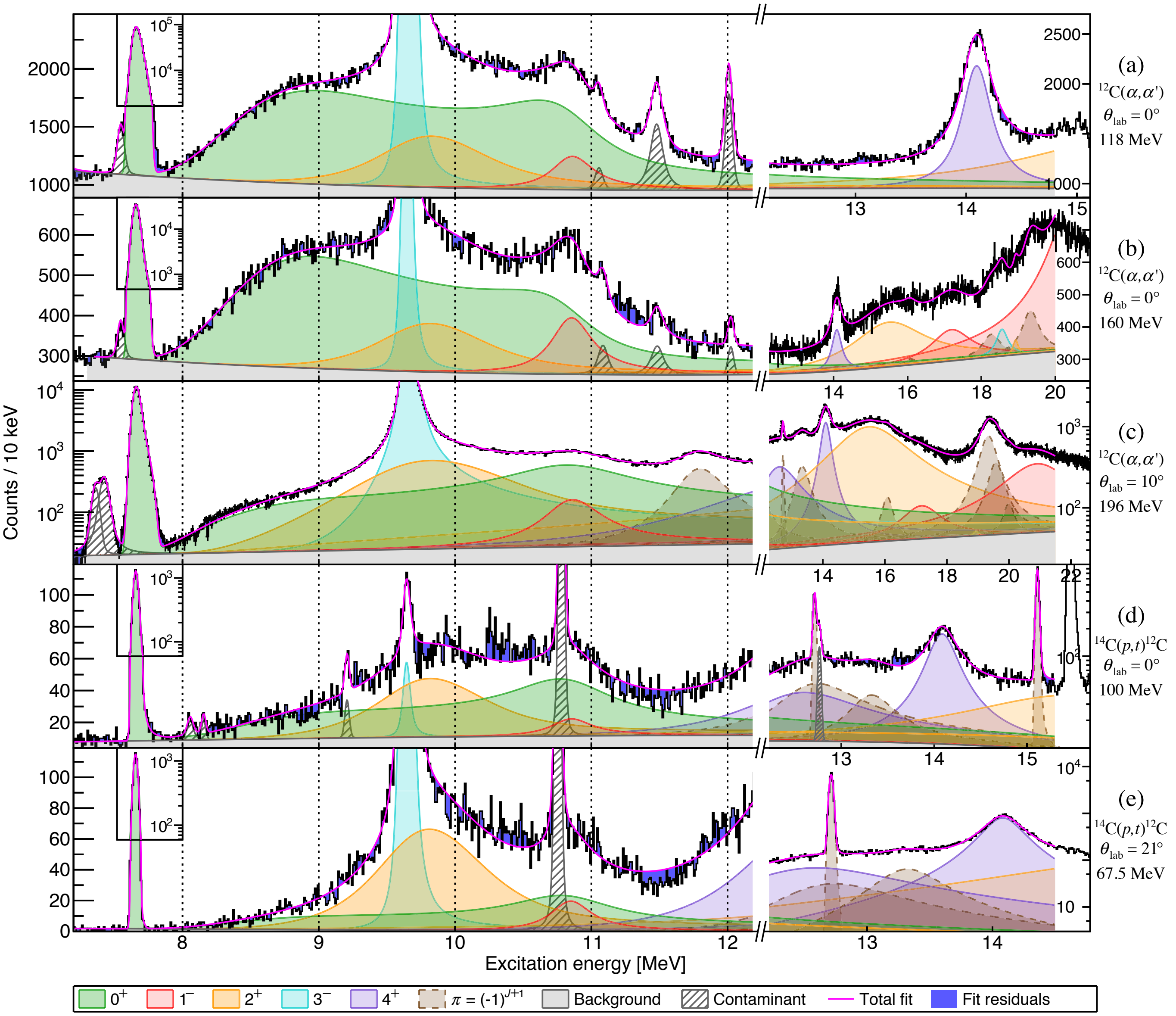

FIG. 15. The optimized global fit for model $M_{0_2^+ 0_\Delta^+ 0_3^+}\left[{}^{-}_{-}\ {}^{+}_{-}\ {}^{-}_{+}\right]$, which accounts for the previously established monopole strengths between $E_x = 7$ and 13 MeV (the $0_2^+$ Hoyle state and a broad $0_3^+$ at $E_x \approx 10$ MeV) and introduces an additional source of monopole strength at $E_x \approx 9$ MeV, denoted $0_\Delta^+$.

**Charged-particle decay:** the relative contributions of different states to the inclusive spectra must match the relative $\ell$-value decompositions of $\alpha_0$ decay to within an order of magnitude (see Section III D 1 and III D 2). The reason for this conservative constraint is because the relative strengths of $\alpha_0$ decay are not directly proportional to the relative inclusive strengths as $\alpha_1$ decay also occurs between $E_x = 7$ and 13 MeV. In this work, $\alpha_1$ decay is not reliably measured in this work due to electronic threshold limitations of the CAKE as well as the difficulty in discriminating the corresponding $\alpha$ particles from $^{12}$C against the $2\alpha$ breakup of $^{8}$Be.

The recommended observable parameters for submodels $M_{0_2^+ 0_\Delta^+ 0_3^+}$ and $M_{0_2^+ 0_\Delta^+ 0_3^+}$ are presented in Tables V and VI, respectively. The decompositions for the average fits of submodels $M_{0_2^+ 0_\Delta^+ 0_3^+}$ and $M_{0_2^+ 0_\Delta^+ 0_3^+}$ are presented in Figs. 17 and 18, respectively. It is observed that the averaged fits of submodels $M_{0_2^+ 0_\Delta^+ 0_3^+}$ and $M_{0_2^+ 0_\Delta^+ 0_3^+}$ both provide good reproductions of the data, with a very low dispersion in the total fits as well as the narrow contam-inant states. Submodel $M_{0^+_2 0^+_\Delta 0^+_3}$ exhibits considerably more variation in the shape and strength of the various contributions in comparison submodel $M_{0^+_2 0^+_\Delta 0^+_3}$ as there are more permutations of interference (and hence, different lineshapes) which qualify for the averaging of submodel $M_{0^+_2 0^+_\Delta 0^+_3}$. The fits with submodels $M_{0^+_2 0^+_\Delta 0^+_3}$ and $M_{0^+_2 0^+_\Delta 0^+_3}$ yield strengths for the $2^+_2$ state populated with both $^{12}$C$(\alpha,\alpha')^{12}$C and $^{14}$C$(p,t)^{12}$C at $\theta_{\rm lab} = 0°$ which are consistent with the charged-particle decay analysis in Sections III D 1 and III D 2, which revealed a considerable $\ell = 2$ contribution between $E_x = 10.0$ to 10.3 MeV. This is a vast improvement over submodel $M_{0^+_2 0^+_3}(+\alpha_0,+\alpha_1)$ which produces a highly suppressed $2^+_2$ state populated with $^{12}$C$(\alpha,\alpha')^{12}$C and $^{14}$C$(p,t)^{12}$C at $\theta_{\rm lab} = 0°$ (see Fig. 13). Finally, submodels $M_{0^+_2 0^+_\Delta 0^+_3}$ and $M_{0^+_2 0^+_\Delta 0^+_3}$ yield broad monopole lineshapes at $E_x \approx 10$ for $^{12}$C$(\alpha,\alpha')^{12}$C at $\theta_{\rm lab} = 0°$ which qualitatively resemble the double-peaked monopole strength from the MDA of Ref. [25] (see Figs. 13 and 14). It was found that all qualifying fits for submodels $M_{0^+_2 0^+_\Delta 0^+_3}$ and $M_{0^+_2 0^+_\Delta 0^+_3}$ correspond to large channel radii of $a_c = 10$ or 11 fm for the additional source of monopole strength $0^+_\Delta$ (out of the discrete set of $a_c = 6.0$ to 11.0 fm in integer steps). This may indicate a spatially extended density and is consistent with the predictions for the breathing-mode excitation of the Hoyle state discussed in Section I and in particular, the proposed channel radius of 10 fm in Ref. [23].

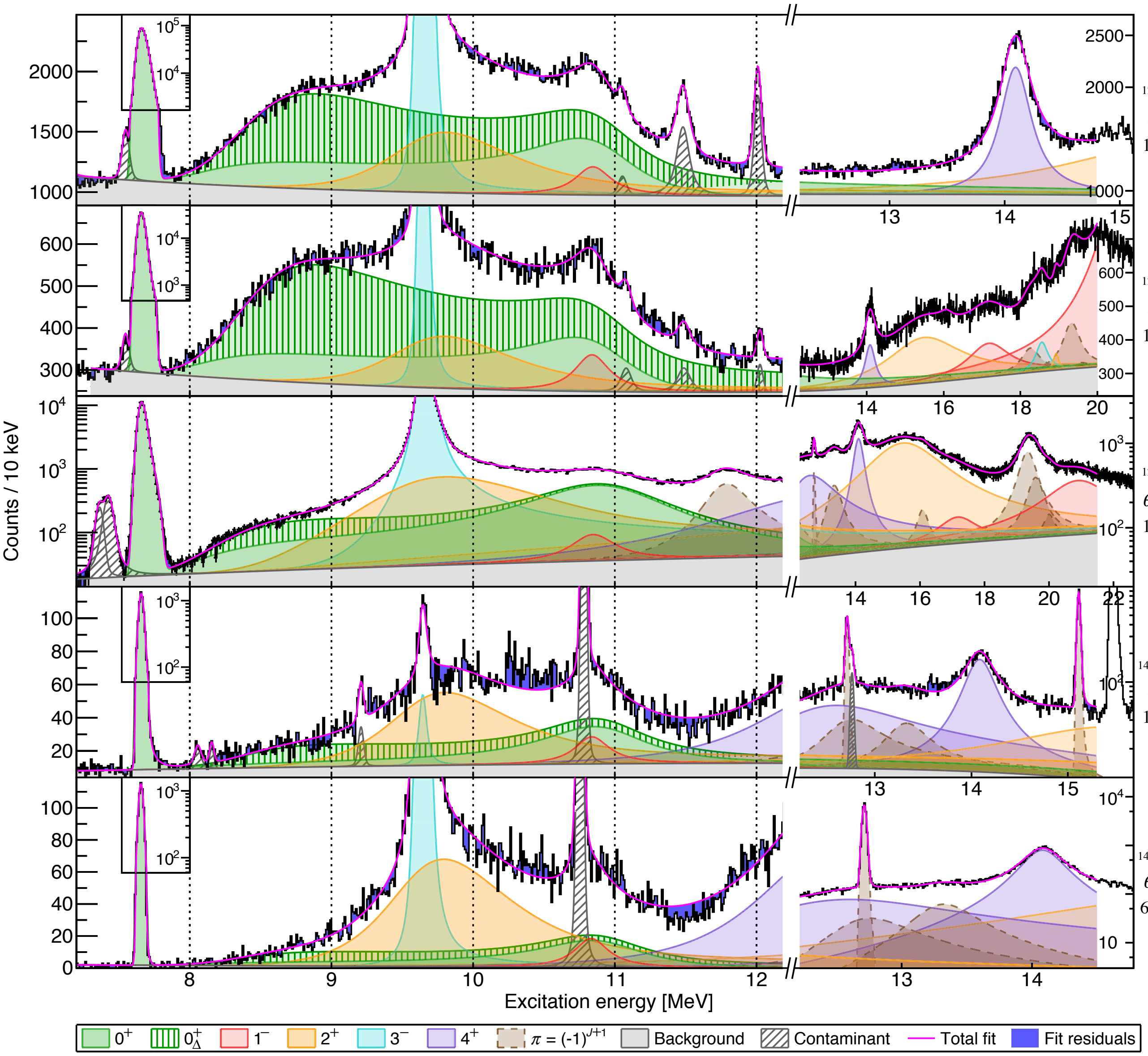

FIG. 16. The optimized global fit for model $M_{0^+_2 0^+_\Delta 0^+_3}(+\alpha_0,-\alpha_1)$, which accounts for the previously established monopole strengths between $E_x = 7$ and 13 MeV (the $0^+_2$ Hoyle state and a broad $0^+_3$ at $E_x \approx 10$ MeV) and introduces an additional source of monopole strength at $E_x \approx 9$ MeV, denoted $0^+_\Delta$. The contribution of the $0^+_\Delta$ response is superimposed on the combined monopole strength from the previously established $0^+_2$ and $0^+_3$ states.

## V. DISCUSSION

### A. Monopole strengths between $E_x = 7$ and 13 MeV

The excitation-energy region of $E_x = 7$ to 13 MeV was studied through both the $^{12}$C$(\alpha, \alpha')^{12}$C and $^{14}$C$(p, t)^{12}$C reactions at various laboratory angles and incident beam energies. The various measurements yielded different relative strengths for the overlapping states between $E_x =$ 7 and 13 MeV, thereby enabling the broad contributions to be disentangled. The $^{12}$C$(\alpha, \alpha')^{12}$C reactions at $\theta_{\rm lab} = 0^\circ$ were observed to strongly populate the collective monopole strengths of the Hoyle state and the broad $0_3^+$ state at $E_x \approx 10$ MeV. This results from the high selectivity for $\alpha$-clustered structures with inelastic alpha scattering as well as the forward-peaked differential cross sections for $0^+$ states. A source of uncertainty in unraveling the broad strengths at $E_x \approx 10$ is the contribution of the $2^+$ rotational excitation of the Hoyle state. In comparison to $^{12}$C$(\alpha, \alpha')^{12}$C at $\theta_{\rm lab} = 0^\circ$, the $^{14}$C$(p, t)^{12}$C reactions (both at $\theta_{\rm lab} = 0^\circ$ and $21^\circ$) employed for this study significantly suppressed the population of the Hoyle state and the broad monopole strength between $E_x = 8$ and 11.5 MeV relative to the $2_2^+$ state. This enabled the intrinsic lineshape of the $2_2^+$ resonance to be accurately parameterized in the global fits, thereby reducing the systematic uncertainty for the contribution of the $2_2^+$ state in the excitation-energy spectra where it was observed to be submerged under broad contributions at $E_x \approx 10$ MeV.

Coincident charged-particle decay measured with the CAKE for the reactions of $^{12}$C$(\alpha, \alpha')^{12}$C and $^{14}$C$(p, t)^{12}$C reactions at $\theta_{\rm lab} = 0^\circ$ enabled the contributions from different states to the broad overlapping structures between $E_x = 7$ and 13 MeV to be disentangled and constrained. The angular correlations indicate the excitation-energy region of $E_x \approx 9$ MeV to be dominantly monopole, which is consistent with the MDA of Ref. [25]. Additionally, the much-debated $2^+$ rotational excitation of the Hoyle state was unequivocally identified, reaffirming previous identifications of this state through both $\alpha$ and proton inelastic scattering as well as photodisintegration [25–27]. Furthermore, the relative strength of the $2_2^+$ state was found to be similar to the sum of other contributions at $E_x \approx 10$ MeV for both the $^{12}$C$(\alpha, \alpha')^{12}$C and $^{14}$C$(p, t)^{12}$C reactions at $\theta_{\rm lab} = 0^\circ$. This enabled the rejection of models which yielded inconsistent relative strengths.

Clear evidence was found in the inclusive measurements for excess monopole strength at $E_x \approx 9$ MeV, which cannot be reproduced by the previously established $0_2^+$ and $0_3^+$ states with the channel-radius dependence and all permutations of constructive/destructive interference being explored. This excess monopole strength cannot be explained by a significantly larger Hoyle state width (currently listed at $\Gamma(E_r) = 9.3(9)$ eV) which alters the contribution from the ghost of the Hoyle state. The introduction of an additional monopole resonance, denoted $0_\Delta^+$, yielded a significantly better reproduction of the data with respect to models which did not include the $0_\Delta^+$ state. Furthermore, the introduction of the $0_\Delta^+$ state yielded relative strengths for the $2_2^+$ state which are consistent with the results of charged-particle decay as well as the MDA of Ref. [25]. In contrast, the models which did not include the additional $0_\Delta^+$ state yielded inconsistent, highly suppressed contributions for the $2_2^+$ state for both the $^{12}$C$(\alpha, \alpha')^{12}$C and $^{14}$C$(p, t)^{12}$C reactions at $\theta_{\rm lab} = 0^\circ$. The highly collective nature of this additional $0_\Delta^+$ state is supported by the fact that a significant excess of monopole strength is not observed at $E_x \approx 9$ MeV in the measurement of $^{14}$C$(p,t)^{12}$C at $\theta_{\rm lab} = 0°$—a reaction that is suggested to suppress collective isoscalar monopole excitations relative to $^{12}$C$(\alpha,\alpha')^{12}$C, because of the different characters of the pairing and radial operators responsible for the two reactions, respectively.

TABLE V. Recommended observable values with monopole interference between the previously established $0_2^+$ (Hoyle) and $0_3^+$ states, as well as the additional $0_\Delta^+$ monopole state, corresponding to model $M_{0_2^+ 0_\Delta^+ 0_3^+}$.

| State | $E_r$ $\pm\sigma_{\rm stat} \pm\sigma_{\rm syst}$ [MeV] | $\Gamma(E_r)$ $\pm\sigma_{\rm stat} \pm\sigma_{\rm syst}$ [keV] | $\Gamma_{\rm FWHM}$ $\pm\sigma_{\rm stat} \pm\sigma_{\rm syst}$ [keV] |
|---|---|---|---|
| $0_2^+$ | 7.65407 fixed | 9.3 fixed | — |
| $0_\Delta^+$ | 9.566 $\pm 0.018 \pm 0.104$ | 3203 $\pm 61 \pm 599$ | — |
| $2_2^+$ | 9.837 $\pm 0.010 \pm 0.079$ | 1181 $\pm 25 \pm 259$ | 921 $\pm 14 \pm 94$ |
| $0_3^+$ | 10.611 $\pm 0.009 \pm 0.309$ | 3473 $\pm 187 \pm 706$ | — |

TABLE VI. Recommended observable values where monopole interference is assumed to only occur between the $0_2^+$ (Hoyle) and $0_3^+$ states. The additional $0_\Delta^+$ monopole state is modelled as an isolated level, corresponding to model $M_{0_2^+ 0_\Delta^+ 0_3^+}$.

| State | $E_r$ $\pm\sigma_{\rm stat} \pm\sigma_{\rm syst}$ [MeV] | $\Gamma(E_r)$ $\pm\sigma_{\rm stat} \pm\sigma_{\rm syst}$ [keV] | $\Gamma_{\rm FWHM}$ $\pm\sigma_{\rm stat} \pm\sigma_{\rm syst}$ [keV] |
|---|---|---|---|
| $0_2^+$ | 7.65407 fixed | 9.3 fixed | — |
| $0_\Delta^+$ | 9.379 $\pm 0.013 \pm 0.050$ | 4565 $\pm 22 \pm 107$ | 2066 $\pm 5 \pm 98$ |
| $2_2^+$ | 9.918 $\pm 0.008 \pm 0.025$ | 1760 $\pm 26 \pm 279$ | 1073 $\pm 6 \pm 54$ |
| $0_3^+$ | 10.969 $\pm 0.007 \pm 0.104$ | 1313 $\pm 25 \pm 395$ | — |

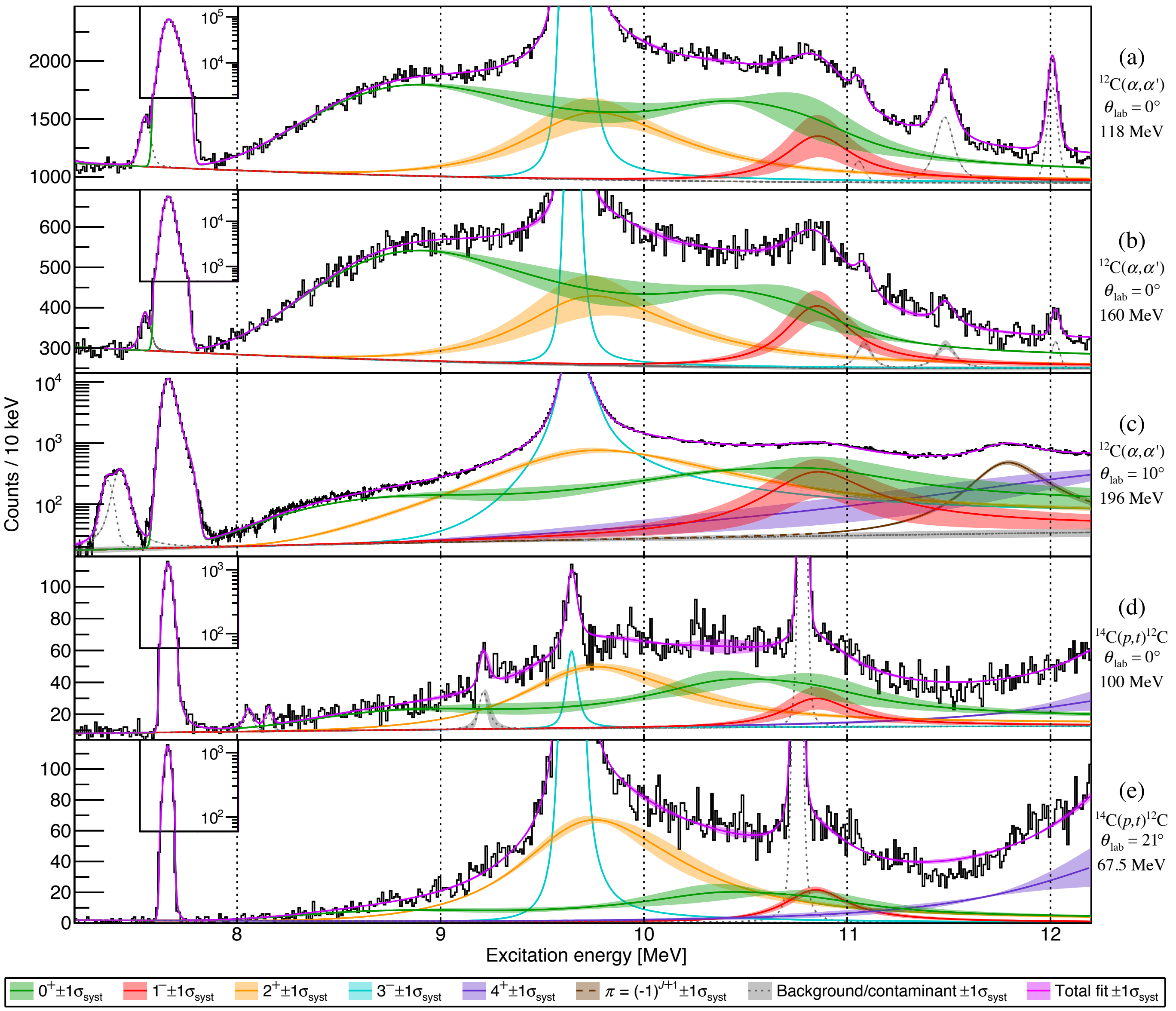

FIG. 17. The average global fit for model $M_{0_2^+ 0_\Delta^+ 0_3^+}\left[\begin{smallmatrix} - & + & - \\ - & - & + \end{smallmatrix}\right]$, which accounts for the previously established monopole strengths between $E_x = 7$ and 13 MeV (the $0_2^+$ Hoyle state and a broad $0_3^+$ at $E_x \approx 10$ MeV) and introduces an additional source of monopole strength at $E_x \approx 9$ MeV, denoted $0_\Delta^+$.

Two forms of parametrization for the monopole strength were investigated: the first being submodel $M_{0_2^+ 0_\Delta^+ 0_3^+}$, whereby the additional $0_\Delta^+$ state was permitted to interfere according to Equation 2 and the recommended parameters are summarized in Table V. For submodel $M_{0_2^+ 0_\Delta^+ 0_3^+}$, the resonance energies of the $0_\Delta^+$ and $0_3^+$ states agree well with theoretical predictions, however the total widths of both states are substantially larger [17–19, 23].

The second form corresponded to submodel $M_{0_2^+ 0_\Delta^+ 0_3^+}$, which treated the additional $0_\Delta^+$ state as an isolated resonance which does not interfere with the previously established $0_2^+$ and $0_3^+$ states. An advantage of this parametrization (if appropriate) is that the FWHM of the intrinsic lineshape for the $0_\Delta^+$ state can be determined. For the submodel $M_{0_2^+ 0_\Delta^+ 0_3^+}$, the recommended parameters are summarized in Table VI. Whilst the $0_\Delta^+$ resonance energy agrees well with Refs. [19, 23], the $\Gamma(E_r)$ total width, is significantly larger than all predictions, similar to submodel $M_{0_2^+ 0_\Delta^+ 0_3^+}$. However, the recommended $\Gamma_{\rm FWHM}$ width is substantially smaller than $\Gamma(E_r)$ and is in good agreement with the predictions of Refs. [18, 19]. It is possible that the $\Gamma(E_r)$ widths for the $0_\Delta^+$ state require more sophisticated models for the $E_x$-dependence in comparison to the **R**-matrix parametrization used in this work.

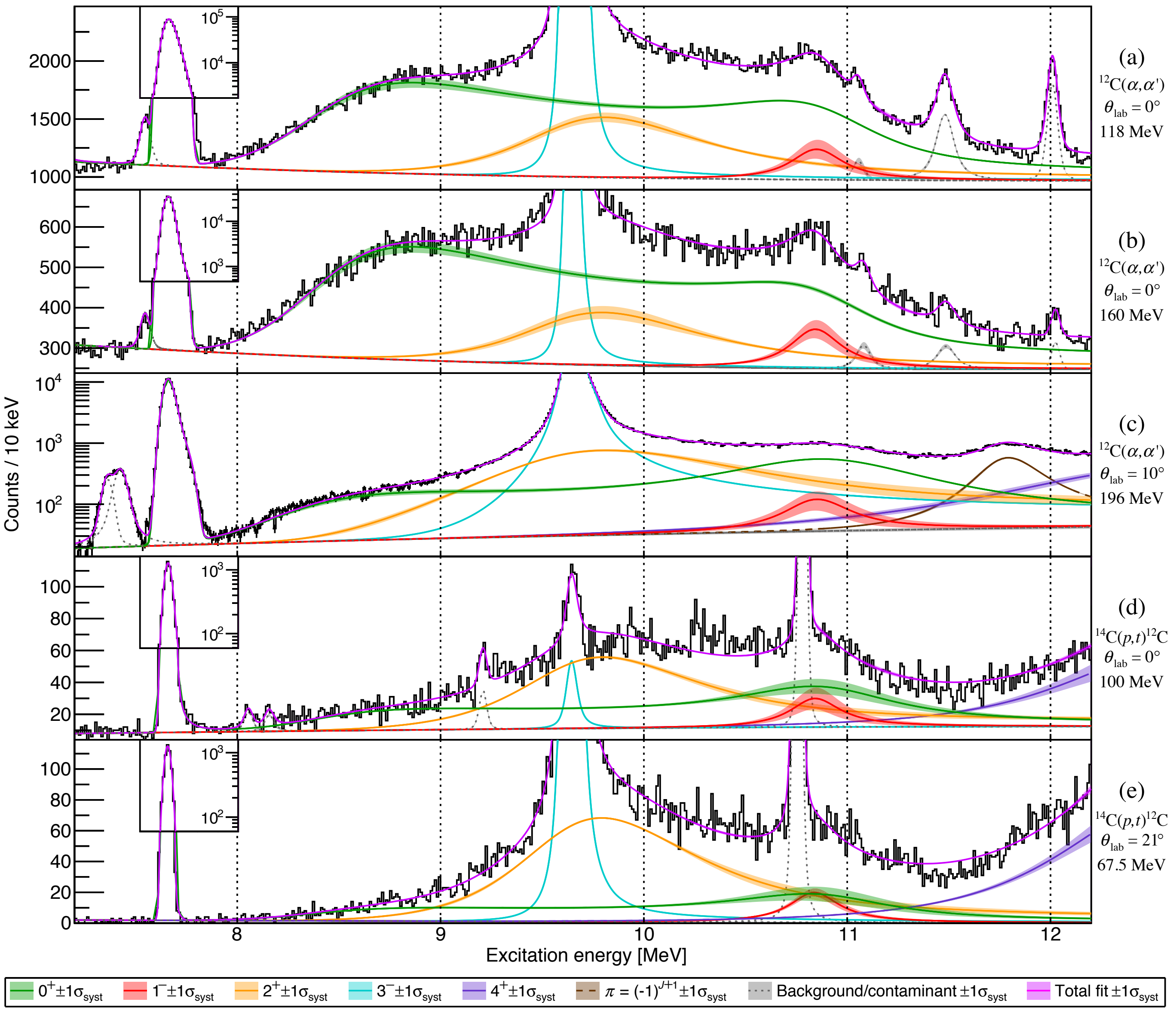

FIG. 18. The average global fit for model $M_{0_2^+ 0_\Delta^+ 0_3^+}(+\alpha_0,-\alpha_1)$, which accounts for the previously established monopole strengths between $E_x = 7$ and 13 MeV (the $0_2^+$ Hoyle state and a broad $0_3^+$ at $E_x \approx 10$ MeV) and introduces an additional source of monopole strength at $E_x \approx 9$ MeV, denoted $0_\Delta^+$.

For the $0_3^+$ state, the recommended parameters for submodel $M_{0_2^+ 0_\Delta^+ 0_3^+}$ are in better agreement with Refs. [18, 19, 23], which predict the $0_3^+$ at $E_x \approx 11$ MeV with $\Gamma \approx 1$ MeV, in comparison to submodel $M_{0_2^+ 0_\Delta^+ 0_3^+}$. The recommended $0_3^+$ parameters for submodel $M_{0_2^+ 0_\Delta^+ 0_3^+}$ are also in good agreement with an **R**-matrix analysis of the $\beta$ decays of $^{12}$N and $^{12}$B which produced $E_x = 11.2(3)$ MeV with $\Gamma = 1.5(6)$ MeV [4]. Interestingly, the beta-decay data in Ref. [4] does not require an additional monopole state $E_x \approx 9$ MeV. This further supports the interpretation of the additional $0_\Delta^+$ state as a breathing-mode excitation composed of coherent 1p-1h excitations as $\beta$-decay is understood to be less selective towards 1p-1h components of collective excitations in comparison to other nuclear/electromagnetic probes [67]. The optimal channel radius for the $0_\Delta^+$ state was found to be $a_c = 11$ fm (out of the discrete set of $a_c = 6.0$ to 11.0 fm in integer steps), which may indicate a spatially extended density and is consistent with the proposed channel radius of 10 fm in Ref. [23].

The location and FWHM of the additional $0^+_\Delta$ state in this work are roughly similar to a Gaussian peak at $E_x \approx 9.04(9)$ MeV with $\Gamma = 1.45(18)$ MeV from a previous study of alpha inelastic scattering in Ref. [25], however there is a statistically significant difference. Furthermore, since the peak-fitting analysis with Gaussian lineshapes in Ref. [25] did not account for the physical effects of near-threshold resonances (such as the "ghost" of the Hoyle state) or interference, care should be taken in this comparison since the two analyses are based on different foundations, resulting in rather different relative population strengths.

It is possible that the excess monopole strength at $E_x \approx 9$ MeV can be alternatively explained by an unsuitable parametrization of how the structure of the Hoyle state evolves with excitation energy. This phenomenological parametrization, which is extensively employed in nuclear astrophysics to extrapolate data to experimentally inaccessible regions [30–35], typically assumes that only two-body effects are significant. Whilst the direct $3\alpha$ decay branch has been shown to be small at the primary peak of the Hoyle state [6–8, 10], the enhanced penetrability at higher excitation-energies may make this direct branch non-negligible. The possibility of such three-body effects is corroborated by the recent indirect measurement of the direct $3\alpha$ decay branch for the Hoyle state, which is predicated on measuring the direct $3\alpha$ decay mode for the $2^+_2$ state at $E_x = 9.6$ MeV, although such decays could not be discriminated from the $\alpha_1$ decay mode [9]. More sophisticated and physically motivated parametrizations of the monopole strength have been reviewed in a recent study [68] and such models should be explored further as inaccurate parametrizations may affect the astrophysical triple-$\alpha$ process, particularly at high temperatures of $T_9 \gtrsim 2$ ($T_9 = T/10^9$ K). Furthermore, the indirect measurement for the total width of the Hoyle state by analyzing the broad monopole strength above the primary peak may be unreliable without appropriate parametrizations. Finally, it is noted that the argument for the existence of a breathing-mode excitation of the Hoyle state is not mutually exclusive with the need for more sophisticated methods of parameterizing both the ghost and the breathing-mode excitation of the Hoyle state.

### B. $2^+$ strength between $E_x = 7$ and 13 MeV

Whilst this work is focused on unraveling the monopole strength of $^{12}$C between $E_x = 7$ and 13 MeV, the $2^+_2$ state requires special attention as for many analyses, its properties are often intertwined with those of the surrounding monopole strength. Currently, the ENSDF database lists the $2^+_2$ state at $E_x = 9.870(60)$ MeV with $\Gamma = 850(85)$ keV. A recent **R**-matrix analysis of photodisintegration data for the $2^+_2$ state produced $E_r = 10.025(50)$ MeV and $\Gamma = 1.60(13)$ MeV [27, 64], which significantly differ from the ENSDF database. One source of this discrepancy is the lack of clarity/consensus in the reporting of widths, e.g. whether a reported width corresponds to $\Gamma(E_r)$ with an **R**-matrix-derived excitation-energy dependence or the FWHM of the intrinsic lineshape. Different analysis methods with varying levels of complexity further compound this uncertainty. To prevent the incorrect limitation of parameters in the fits of this work, the $2^+_2$ parameters were not constrained purely with the ENSDF-listed values. For example, the total width was constrained by an upper limit of $3\sigma$ above $\Gamma = 1.60(13)$ MeV corresponding to Ref. [64]. Since photodisintegration does not populate the surrounding broad monopole strength, the $2^+_2$ parameters from Ref. [64] were assumed to be more reliable.

For submodels $M_{0^+_2\,0^+_\Delta\,0^+_3}$ and $M_{0^+_2\,0^+_\Delta\,0^+_3}$, the $2^+_2$ state parameters are summarized in Tables V and VI, respectively. Both of these submodels provide similar descriptions of the data, yielding resonance energies which are in relatively good agreement with one another and are both within $1\sigma$ of the ENSDF-listed value of $E_x = 9.870(60)$ MeV. The recommended $\Gamma(E_r)$ widths from this work are larger than the ENSDF-listed value of $E_x = 850(85)$ keV, with the width of $\Gamma(E_r) = 1.760 \pm 0.026_{(\mathrm{stat})} \pm 0.279_{(\mathrm{syst})}$ MeV in particularly good agreement with the width of $\Gamma = 1.60(13)$ MeV from the photodisintegration data [64].

The recommended $\Gamma_{\mathrm{FWHM}}$ widths for the $2^+_2$ state from this work are significantly smaller than the corresponding $\Gamma(E_r)$ total widths evaluated at the resonance energy and are in better agreement with the ENSDF-listed width of $\Gamma = 850(85)$ keV. This substantial difference between $\Gamma(E_r)$ and $\Gamma_{\mathrm{FWHM}}$ occurs for resonances with broad widths near the corresponding particle thresholds. This results in a lineshape with a low-$E_x$ tail which is highly suppressed by the diminishing penetrability.

In this work, there is no evidence for two distinct $2^+$ states between $E_x = 7$ and 13 MeV. This corroborates a study using the $^{11}$B($^3$He, $d$)$^{12}$C reaction [69] which found no evidence for a $2^+$ state previously suggested at $E_x = 11.16$ MeV [70]. Whilst Ref. [27] makes the suggestion of an additional $2^+$ state due to a single, inconsistent data point, additional supporting evidence with a higher-$E_x$ range is required to confirm this hypothesis. If two distinct $2^+$ states were to exist between $E_x = 7$ and 13 MeV, it would be expected that the various reactions employed in this work would produce different relative populations between the two distinct $2^+$ states. In such a case, the simultaneous fits of the inclusive spectra in this work should identify models with a single $2^+$ state between $E_x = 7$ and 13 MeV as inconsistent with the data.

## VI. CONCLUSION

Knowledge of the low-lying monopole strength in $^{12}$C—the Hoyle state in particular—is crucial for our understanding of both the astrophysically important $3\alpha$ reaction and of $\alpha$-particle clustering. Recent theoretical calculations predict a breathing-mode excitation of the Hoyle state at $E_x \approx 9$ MeV with a width of $\Gamma \approx 1.5$ MeV. The observation of this breathing-mode excitation is hindered by the presence of multiple broad states and potential interference effects. The $^{12}\mathrm{C}(\alpha,\alpha')^{12}\mathrm{C}$ and $^{14}\mathrm{C}(p,t)^{12}\mathrm{C}$ reactions were employed to populate states in $^{12}$C. A self-consistent, simultaneous analysis of the inclusive spectra with lineshapes accounting for experimental effects and distortion due to nuclear dynamics yielded clear evidence for excess monopole strength at $E_x \approx 9$ MeV, particularly for $^{12}\mathrm{C}(\alpha,\alpha')^{12}\mathrm{C}$ at $\theta_{\mathrm{lab}} = 0°$ and the data is not well reproduced by the previously established $0_2^+$ and $0_3^+$ states. The analysis of coincident charged-particle decay data supports this conclusion. An additional monopole state at $E_x \approx 9$ MeV, denoted $0_\Delta^+$, significantly improved the description of both inclusive and charged-particle-gated data. This new monopole state is the leading candidate for the breathing-mode excitation of the Hoyle state and two parametrizations for the sources of monopole strength between $E_x = 7$ and 13 MeV were employed. The first being submodel $M_{0_2^+ 0_\Delta^+ 0_3^+}$, whereby the three sources of monopole strength ($0_2^+$, $0_\Delta^+$ and $0_3^+$) were permitted to interfere according to Equation 2. For this submodel, the $0_\Delta^+$ was optimized at $E_x = 9.566 \pm 0.018_{(\mathrm{stat})} \pm 0.104_{(\mathrm{syst})}$ MeV with $\Gamma(E_r) = 3.203 \pm 0.061_{(\mathrm{stat})} \pm 0.599_{(\mathrm{syst})}$ MeV (see Table V for details). The second parametrization of the monopole strengths corresponded to submodel $M_{0_2^+ 0_\Delta^+ 0_3^+}$, whereby the additional $0_\Delta^+$ state was treated as an isolated resonance whilst the previously established $0_2^+$ and $0_3^+$ states were permitted to interfere. This parametrization may be appropriate if the $0_\Delta^+$ state corresponds to a dynamic, collective monopole response which does not interfere with the neighboring $0^+$ states which are understood to be statically deformed. This submodel yielded $E_x = 9.379 \pm 0.013_{(\mathrm{stat})} \pm 0.050_{(\mathrm{syst})}$ MeV with $\Gamma(E_r) = 4.565 \pm 0.022_{(\mathrm{stat})} \pm 0.107_{(\mathrm{syst})}$ MeV with $\Gamma_{\mathrm{FWHM}} = 2.066 \pm 0.005_{(\mathrm{stat})} \pm 0.098_{(\mathrm{syst})}$ MeV. For the $\Gamma(E_r)$ width of the $0_\Delta^+$ state, better agreement may be achieved with theoretical predictions by using more sophisticated models for the excitation-energy dependence compared to the **R**-matrix parametrization used in this work. In contrast, the recommended $\Gamma_{\mathrm{FWHM}}$ width may be more model independent than $\Gamma(E_r)$ and is in good agreement with the predictions of Refs. [18, 19].

The highly collective nature of this additional $0_\Delta^+$ state is supported by the fact that a significant excess of monopole strength is not observed at $E_x \approx 9$ MeV in the measurement of $^{14}\mathrm{C}(p,t)^{12}\mathrm{C}$ at $\theta_{\mathrm{lab}} = 0°$—a reaction that is suggested to suppress collective isoscalar monopole excitations relative to $^{12}\mathrm{C}(\alpha,\alpha')^{12}\mathrm{C}$, because of the different characters of the pairing and radial operators responsible for the two reactions, respectively. The interpretation of the $0_\Delta^+$ state as a breathing-mode excitation, composed of coherent 1p-1h excitations, is further supported by its weak population through $\beta$-decay [4], which is understood to be less selective towards 1p-1h components of collective excitations in comparison to other nuclear/electromagnetic probes [67]. An alternative explanation, which must be considered, is that the excess monopole strength is symptomatic of a requirement for more sophisticated theoretical descriptions of the properties of the Hoyle state, which may influence the temperature-dependence of the $3\alpha$ rate at $T_9 \gtrsim 2$.

## ACKNOWLEDGMENTS

This work is based on the research supported in part by the National Research Foundation of South Africa (Grant Numbers: 85509, 86052, 118846, 90741). The authors acknowledge the accelerator staff of iThemba LABS for providing excellent beams. The authors would like to thank E. Khan, A. Tamii, B. Zhou, K. Masaaki, H.O.U. Fynbo, M. Itoh and J. Carter for useful discussions. PA acknowledges support from the Claude Leon Foundation in the form of a postdoctoral fellowship. The computations were performed on resources provided by UNINETT Sigma2 - the National Infrastructure for High Performance Computing and Data Storage in Norway. The authors are grateful to A.C. Larsen, F. Zeiser and F. Pogliano for their assistance with UNINETT Sigma2.

### Appendix: Fit results employing the alternative penetrability prescription of Equation 6

In Section IV, the presented results correspond to the penetrability prescription of Equation 5. In this section, a set of analogous results are presented using the alternative penetrability prescription of Equation 6. The fit results presented in this section are highly similar to those in Section IV and lead to the same conclusions. Consequently, the explanations for the fit results in Section IV are not repeated in this section.

The optimal fits for submodels $M_{0_2^+}$, $M_{0_2^+ 0_3^+(+\alpha_0,-\alpha_1)}$, $M_{0_2^+ 0_\Delta^+ 0_3^+}\left[\begin{smallmatrix} - & + & - \\ - & - & - \end{smallmatrix}\right]$ and $M_{0_2^+ 0_\Delta^+ 0_3^+(+\alpha_0,-\alpha_1)}$ are presented in Figs. 19, 20, 21 and 22, respectively. The corresponding fit results are summarized in Table VII.

The recommended observable parameters for submodel $M_{0_2^+ 0_\Delta^+ 0_3^+}\left[\begin{smallmatrix} - & + & - \\ - & - & - \end{smallmatrix}\right]$ are summarized in Table VIII with the corresponding decomposition in Fig. 23. The recommended observable parameters for $M_{0_2^+ 0_\Delta^+ 0_3^+(+\alpha_0,-\alpha_1)}$ are summarized in Tables IX with the corresponding decomposition in Fig. 24. It is observed that the observable parameters extracted using the penetrability prescription of Equation 6 agree well with the results in Section IV which correspond to the penetrability prescription of Equation 5.

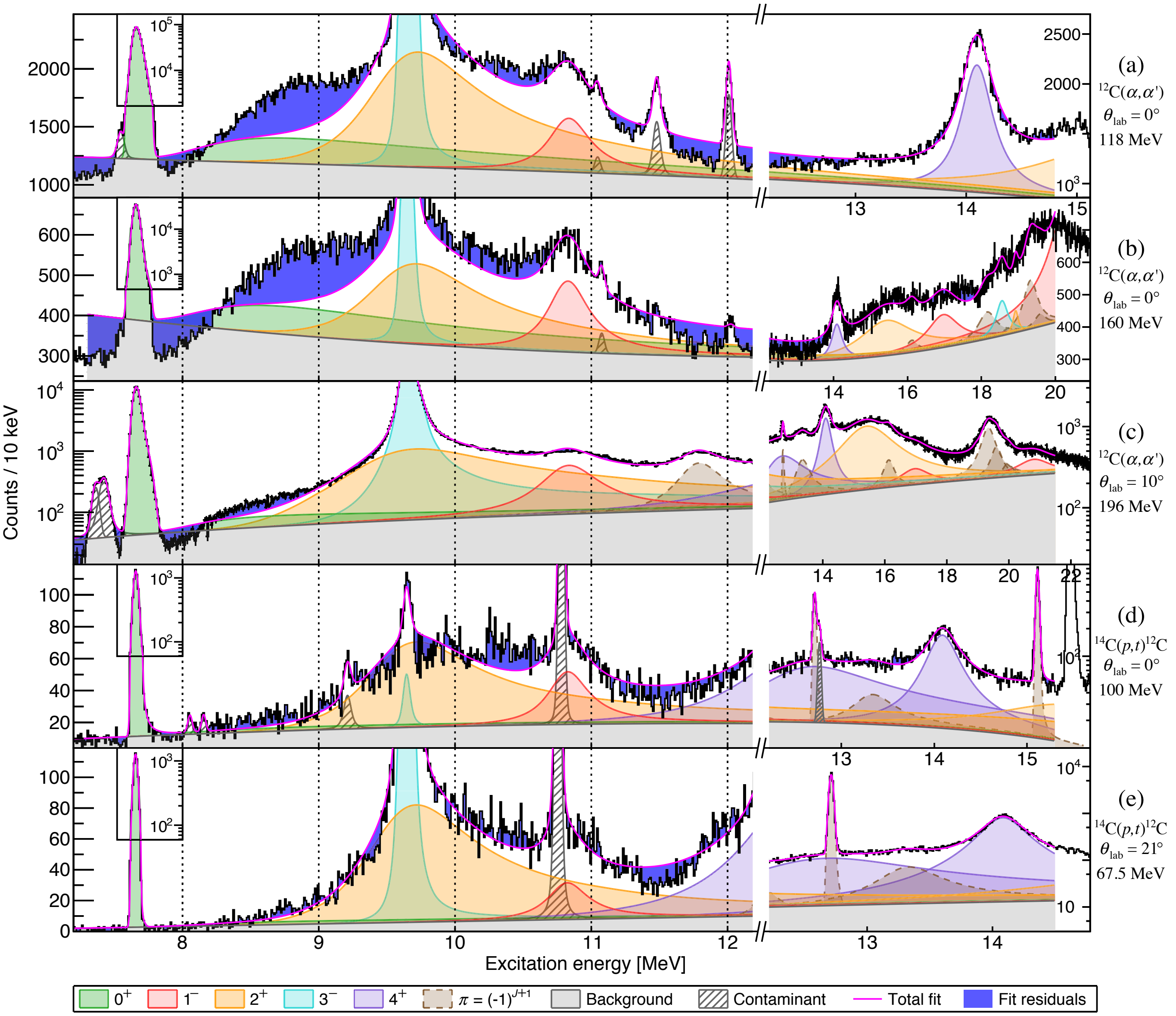

FIG. 19. The optimized global fit for model $M_{0_2^+}$(performed with the penetrability prescription of Equation 6): it is assumed that the broad monopole strength above the primary peak of the $0_2^+$ Hoyle state and below $E_x \approx 13$ MeV corresponds only to the ghost of the Hoyle state. See appendix for details.

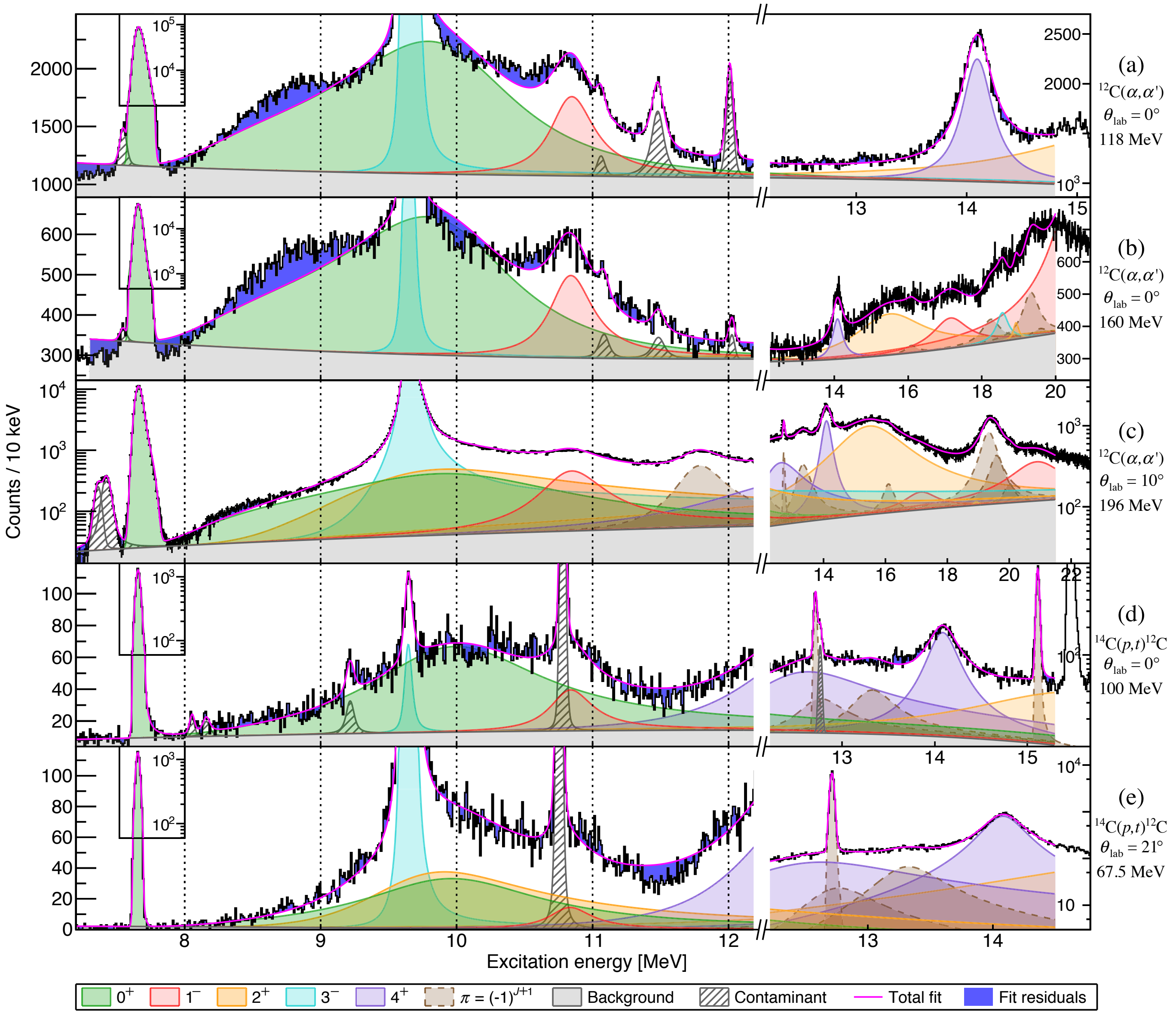

FIG. 20. The optimized global fit for submodel $M_{0_2^+ 0_3^+(+\alpha_0,-\alpha_1)}$(performed with the penetrability prescription of Equation 6), which accounts for all of the previously established sources of monopole strength between $E_x = 7$ and 13 MeV: the $0_2^+$ Hoyle state and a broad $0_3^+$ resonance at $E_x \approx 10$ MeV. The results of this fit are summarized in Table VII. See appendix for details.

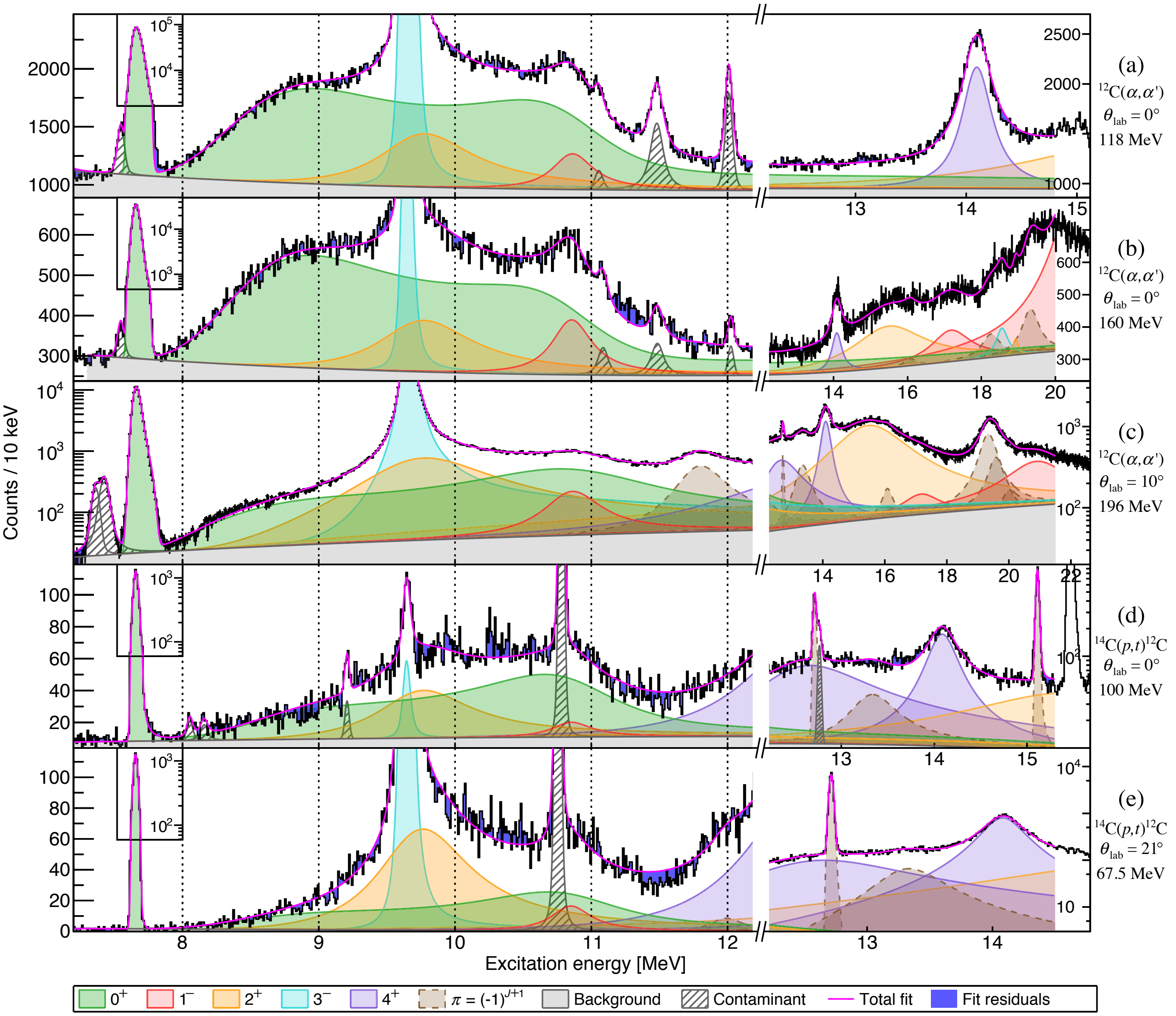

FIG. 21. The optimized global fit for model $M_{0_2^+\,0_\Delta^+\,0_3^+}\left[\begin{smallmatrix} - & + & - \\ - & - & - \end{smallmatrix}\right]$(performed with the penetrability prescription of Equation 6), which accounts for the previously established monopole strengths between $E_x = 7$ and 13 MeV (the $0_2^+$ Hoyle state and a broad $0_3^+$ at $E_x \approx 10$ MeV) and introduces an additional source of monopole strength at $E_x \approx 9$ MeV, denoted $0_\Delta^+$.

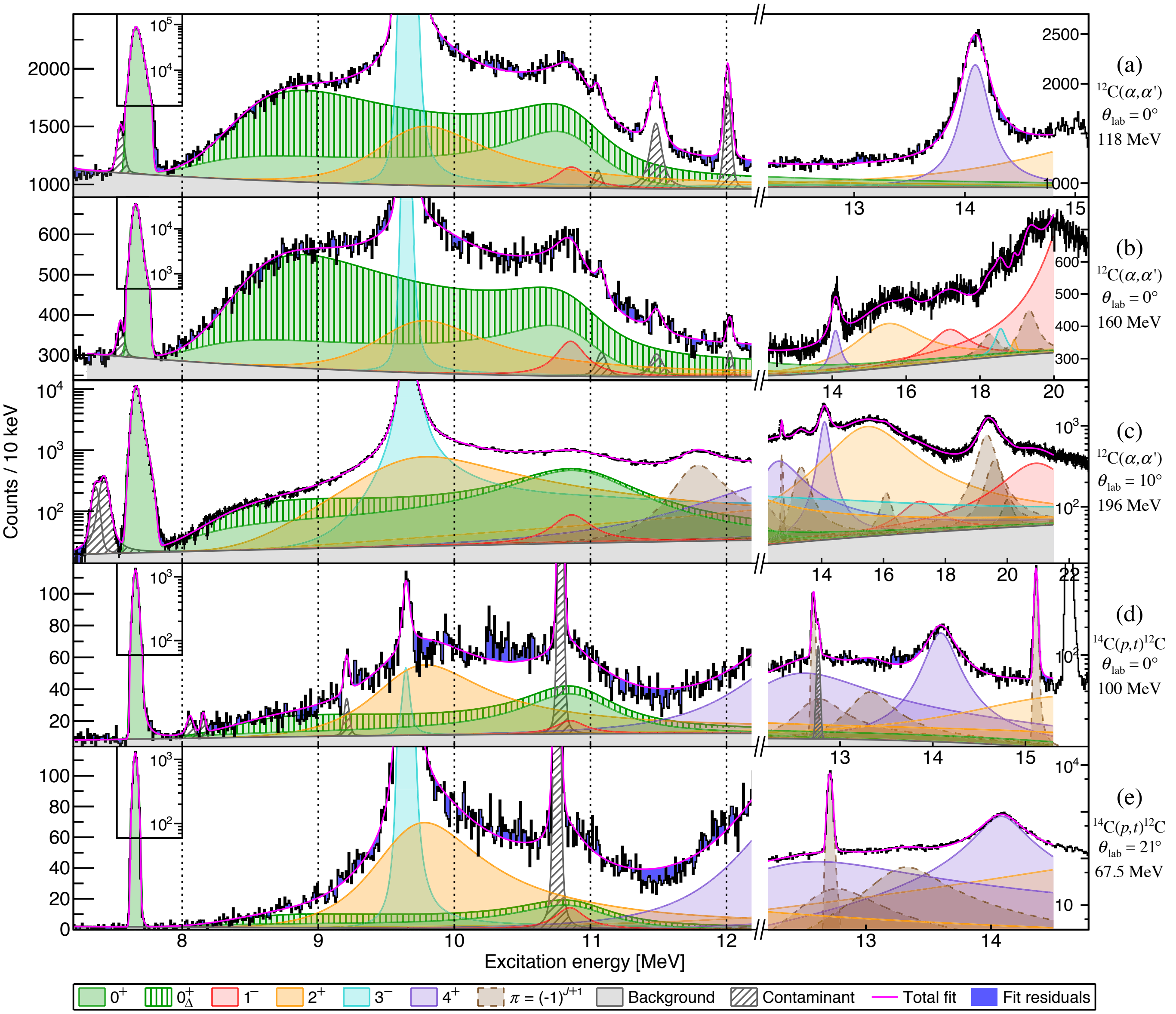

FIG. 22. The optimized global fit for model $M_{0_2^+\,0_\Delta^+\,0_3^+}(+\alpha_0,-\alpha_1)$(performed with the penetrability prescription of Equation 6), which accounts for the previously established monopole strengths between $E_x = 7$ and 13 MeV (the $0_2^+$ Hoyle state and a broad $0_3^+$ at $E_x \approx 10$ MeV) and introduces an additional source of monopole strength at $E_x \approx 9$ MeV, denoted $0_\Delta^+$. The contribution of the $0_\Delta^+$ response is superimposed on the combined monopole strength from the previously established $0_2^+$ and $0_3^+$ states.

TABLE VII. Summary of the optimal fit results with the penetrability prescription of Equation 6. Errors for the resonance energies account for both the fit error and an estimated focal-plane calibration error whilst the errors for the widths are purely from the fit.

| Model | AIC<br>BIC | State | $E_r$<br>[MeV] | $\Gamma_{\alpha_0}(E_r)$<br>[keV] | $\theta^2_{\alpha_0}$ | $r_{\alpha_0}$<br>[fm] | $\Gamma_{\alpha_1}(E_r)$<br>[keV] | $\theta^2_{\alpha_1}$ | $r_{\alpha_1}$<br>[fm] | $\Gamma(E_r)$<br>[keV] | $\Gamma_{\mathrm{FWHM}}$<br>[keV] |
|---|---|---|---|---|---|---|---|---|---|---|---|
| $M_{0_2^+}$ | 29334 | $0_2^+$ | [a]7.65407 | [a]$9.3\times10^{-3}$ | 0.50 | 7 | $\approx 0$ | $\approx 0$ | 7 | $9.3\times10^{-3}$ | |
| | 28984 | $2_2^+$ | 9.868(7) | 1809(18) | 1.302(13) | 7 | 10.3(1) | 2.09(1) | 7 | 1819(18) | 1127(5) |
| $M_{0_2^+0_3^+}$ | 15855 | $0_2^+$ | [a]7.65407 | [a]$9.3\times10^{-3}$ | 0.28 | 8 | $\approx 0$ | $\approx 0$ | 8 | $9.3\times10^{-3}$ | |
| $\left[\begin{smallmatrix}+&-\\+&-\end{smallmatrix}\right]$ | 15489 | $2_2^+$ | 10.143(21) | 2324(17) | 1.34(1) | 8 | $\approx 0$ | $\approx 0$ | 8 | 2324(17) | 1664(5) |
| | | $0_3^+$ | 10.282(11) | $\approx 2367$ | $\approx 1$ | 8 | $\approx 0$ | $\approx 0$ | 8 | $\approx 2367$ | |
| $M_{0_2^+0_\Delta^+0_3^+}$ | 12767 | $0_2^+$ | [a]7.65407 | [a]$9.3\times10^{-3}$ | 0.095 | 10 | $\approx 0$ | $\approx 0$ | 10 | $9.3\times10^{-3}$ | |
| | 12385 | $0_\Delta^+$ | 9.624(31) | 3611(42) | 2.224(26) | 10 | $\approx 0$ | $\approx 0$ | 10 | 3611(42) | |
| $\left[\begin{smallmatrix}-&+&-\\-&-&-\end{smallmatrix}\right]$ | | $2_2^+$ | 9.815(15) | 903(48) | 0.615(33) | 10 | 3(22) | 0.4(31) | 10 | 905(53) | 837(49) |
| | | $0_3^+$ | 10.919(19) | 1310(86) | 0.545(36) | 10 | 13(9) | 0.14(9) | 10 | 1323(86) | |
| $M_{0_2^+0_\Delta^+0_3^+}$ | 12828 | $0_2^+$ | [a]7.65407 | [a]$9.3\times10^{-3}$ | 0.28 | 8 | $\approx 0$ | $\approx 0$ | 8 | $9.3\times10^{-3}$ | |
| | 12446 | $0_\Delta^+$ | 9.468(21) | 4316(84) | 3.01(6) | 11 | 8(1) | 5.1(4) | 11 | 4323(84) | 2184(1) |
| $\left[\begin{smallmatrix}+&+\\-&-\end{smallmatrix}\right]$ | | $2_2^+$ | 9.888(10) | 1335(23) | 0.886(15) | 8 | 7(5) | 1.03(68) | 8 | 1341(23) | 1064(12) |
| | | $0_3^+$ | 10.968(10) | 1020(28) | 0.356(10) | 8 | $\approx 0$ | $\approx 0$ | 8 | 1020(28) | |

[a] Fixed parameters in the fit optimisation.

TABLE VIII. Recommended observable values with monopole interference between the previously established $0_2^+$ (Hoyle) and $0_3^+$ states, as well as the additional $0_\Delta^+$ monopole state, corresponding to model $M_{0_2^+0_\Delta^+0_3^+}$.

| State | $E_r$<br>$\pm\sigma_{\mathrm{stat}}\pm\sigma_{\mathrm{syst}}$<br>[MeV] | $\Gamma(E_r)$<br>$\pm\sigma_{\mathrm{stat}}\pm\sigma_{\mathrm{syst}}$<br>[keV] | $\Gamma_{\mathrm{FWHM}}$<br>$\pm\sigma_{\mathrm{stat}}\pm\sigma_{\mathrm{syst}}$<br>[keV] |
|---|---|---|---|
| $0_2^+$ | 7.65407<br>fixed | 9.3<br>fixed | — |
| $0_\Delta^+$ | 9.664<br>$\pm 0.015 \pm 0.070$ | 3596<br>$\pm 29 \pm 78$ | — |
| $2_2^+$ | 9.830<br>$\pm 0.008 \pm 0.032$ | 981<br>$\pm 19 \pm 53$ | 893<br>$\pm 16 \pm 53$ |
| $0_3^+$ | 10.762<br>$\pm 0.014 \pm 0.351$ | 1814<br>$\pm 86 \pm 787$ | — |

TABLE IX. Recommended observable values where monopole interference is assumed to only occur between the $0_2^+$ (Hoyle) and $0_3^+$ states. The additional $0_\Delta^+$ monopole state is modelled as an isolated level, corresponding to model $M_{0_2^+0_\Delta^+0_3^+}$.

| State | $E_r$<br>$\pm\sigma_{\mathrm{stat}}\pm\sigma_{\mathrm{syst}}$<br>[MeV] | $\Gamma(E_r)$<br>$\pm\sigma_{\mathrm{stat}}\pm\sigma_{\mathrm{syst}}$<br>[keV] | $\Gamma_{\mathrm{FWHM}}$<br>$\pm\sigma_{\mathrm{stat}}\pm\sigma_{\mathrm{syst}}$<br>[keV] |
|---|---|---|---|
| $0_2^+$ | 7.65407<br>fixed | 9.3<br>fixed | — |
| $0_\Delta^+$ | 9.438<br>$\pm 0.018 \pm 0.027$ | 4441<br>$\pm 61 \pm 285$ | 2105<br>$\pm 2 \pm 70$ |
| $2_2^+$ | 9.890<br>$\pm 0.008 \pm 0.007$ | 1425<br>$\pm 19 \pm 161$ | 1067<br>$\pm 8 \pm 13$ |
| $0_3^+$ | 10.990<br>$\pm 0.007 \pm 0.042$ | 1196<br>$\pm 25 \pm 240$ | — |

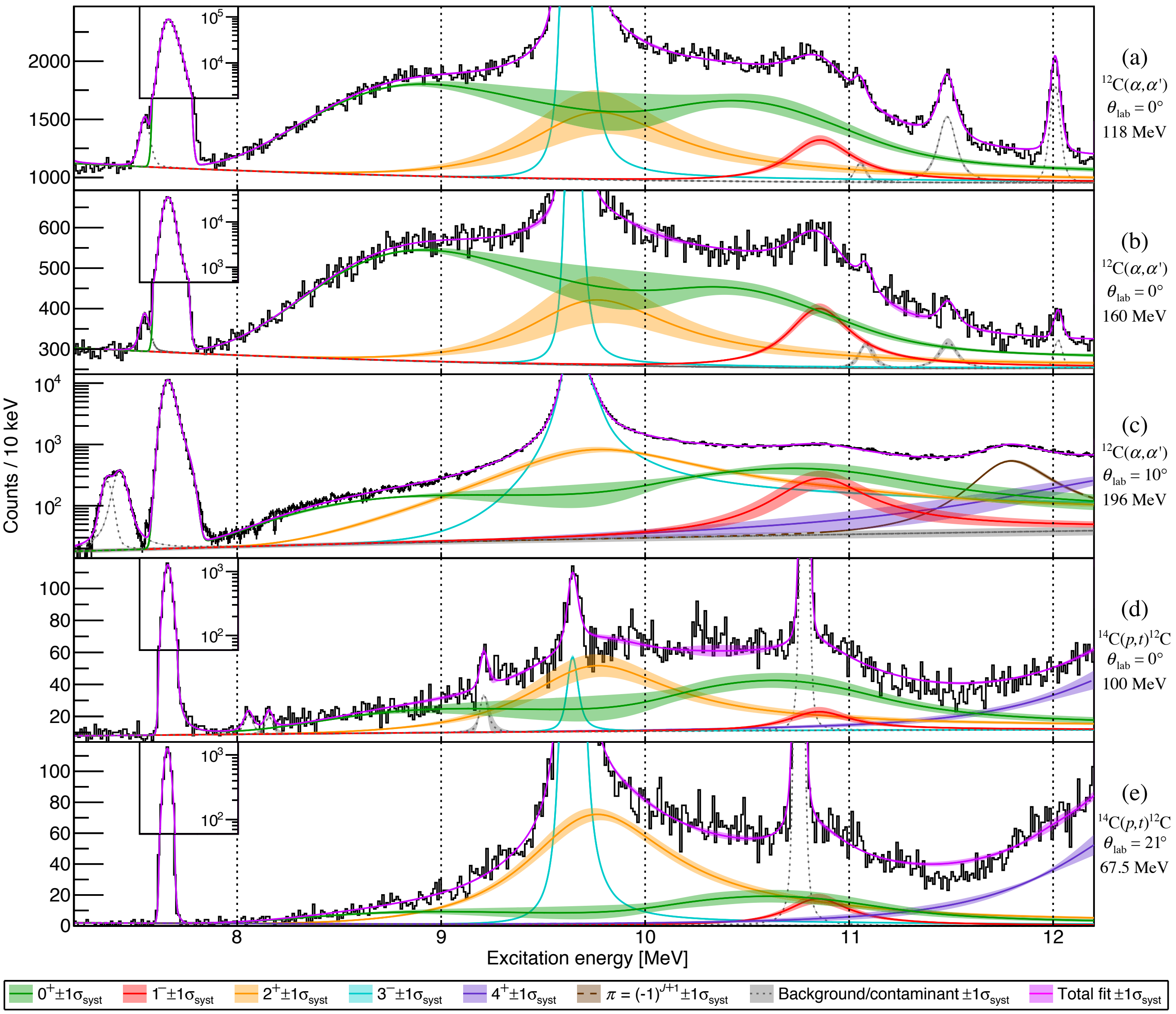

FIG. 23. The averaged global fit for model $M_{0_2^+ 0_\Delta^+ 0_3^+}\left[\begin{smallmatrix} - & + & - \\ - & - & + \end{smallmatrix}\right]$, which accounts for the previously established monopole strengths between $E_x = 7$ and 13 MeV (the $0_2^+$ Hoyle state and a broad $0_3^+$ at $E_x \approx 10$ MeV) and introduces an additional source of monopole strength at $E_x \approx 9$ MeV, denoted $0_\Delta^+$.

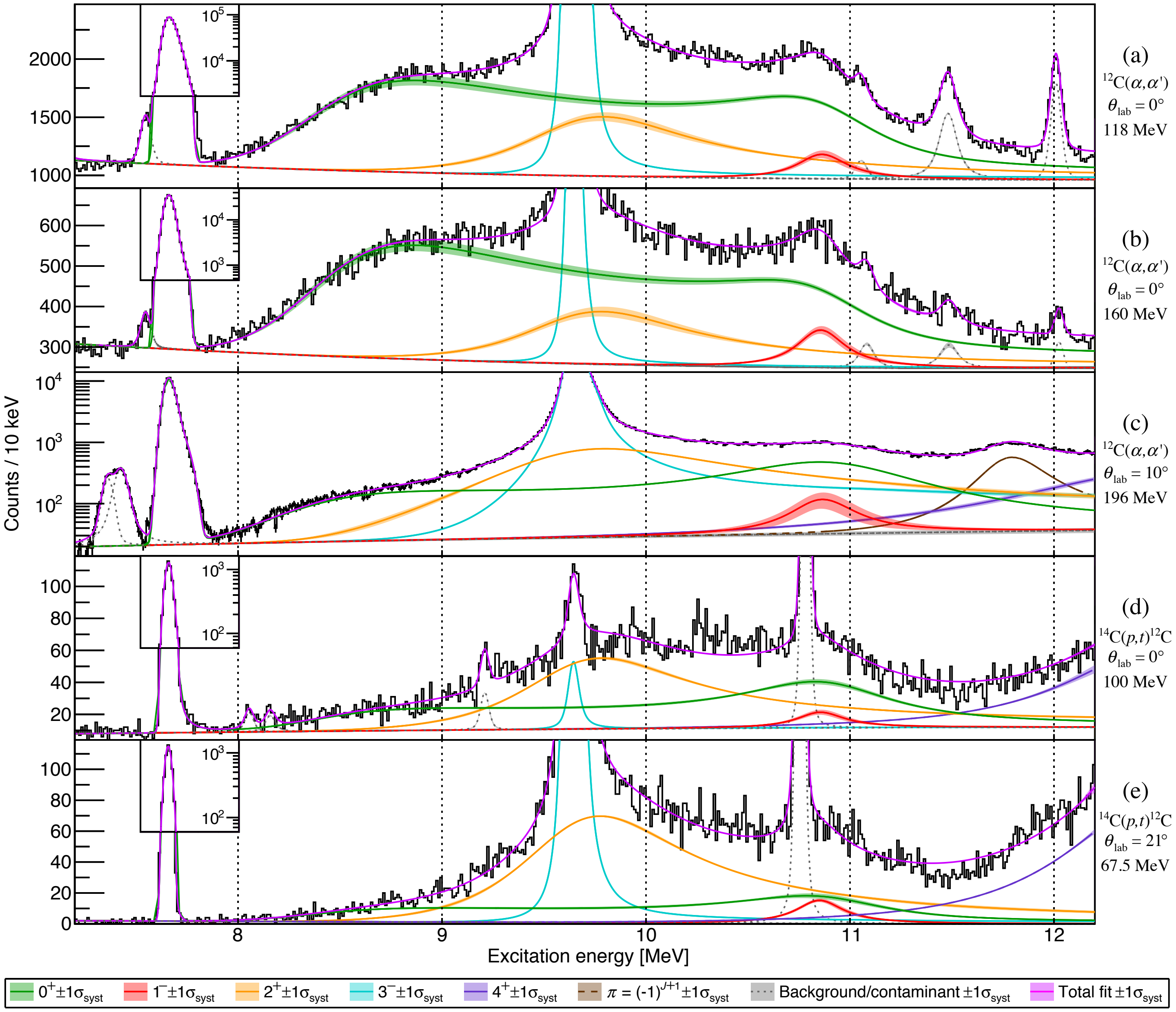

FIG. 24. The averaged global fit for model $M_{0_2^+\,0_\Delta^+\,0_3^+}(+\alpha_0,-\alpha_1)$, which accounts for the previously established monopole strengths between $E_x = 7$ and 13 MeV (the $0_2^+$ Hoyle state and a broad $0_3^+$ at $E_x \approx 10$ MeV) and introduces an additional source of monopole strength at $E_x \approx 9$ MeV, denoted $0_\Delta^+$.